\documentclass[twocolumn]{aastex701}

\usepackage{amsmath}

\begin{document}

\title{Mid-Infrared Diagnostics of PAH Processing and Radiation Fields in M33 H II Regions with JWST/MIRI}

\author[orcid=0009-0002-6147-531X,sname='Porel']{Puja Porel}
\affiliation{Indian Institute of Astrophysics, II Block, Koramangala, Bengaluru 560034, India}
\affiliation{Pondicherry University, R.V. Nagar, Kalapet, 605014, Puducherry, India}
\email[show]{pujaporel11@gmail.com}  

\author[orcid=0000-0002-6386-2906, sname='Soam']{Archana Soam} 
\affiliation{Indian Institute of Astrophysics, II Block, Koramangala, Bengaluru 560034, India}
\affiliation{Pondicherry University, R.V. Nagar, Kalapet, 605014, Puducherry, India}
\email{archana.soam@iiap.res.in}

\author[orcid=0009-0002-6171-9740, sname='Saikhom']{Saikhom Pravash} 
\affiliation{Indian Institute of Astrophysics, II Block, Koramangala, Bengaluru 560034, India}
\affiliation{Pondicherry University, R.V. Nagar, Kalapet, 605014, Puducherry, India}
\email{archana.soam@iiap.res.in}


\begin{abstract}

We present \textit{JWST}/MIRI (Proposal ID 4297) medium-resolution integral field spectroscopy of five H\,\textsc{ii} regions in the nearby spiral galaxy M33, obtained from the Mikulski Archive for Space Telescopes (MAST). The fully calibrated Level-3 spectral cubes provide continuous wavelength coverage from 6.5--20.9~$\mu$m, enabling detailed analysis of polycyclic aromatic hydrocarbon (PAH) emission, ionic fine-structure lines, and warm molecular hydrogen. The spectra exhibit prominent PAH bands at 7.7, 8.6, and 11.3~$\mu$m, multiple H$_2$ pure rotational transitions, and hydrogen recombination lines. From H$_2$ rotational diagrams, we derive warm molecular gas temperatures of $\sim$316--407~K. The detection of [Ne\,\textsc{ii}], [Ne\,\textsc{iii}], [S\,\textsc{iii}], and [S\,\textsc{iv}] enables measurements of ionic abundances and radiation-field hardness, yielding ionization indices of $-1.41$ to $+0.98$. We find that the PAH ionization fraction shows no clear dependence on radiation hardness, consistent with previous \textit{Spitzer}-based results. However, unlike previous work, we find no clear correlation between the 7.7/11.2 and 8.6/11.2~$\mu$m PAH ratios and a moderate negative correlation between 7.7/11.2~$\mu$m and metallicity, possibly due to our limited sample size ($N=5$). The PAH-to-VSG ratio decreases with increasing radiation hardness. We derive the ionization parameter $\gamma$, far-ultraviolet radiation field, and PAH charge state, finding a predominantly cationic PAH population ($\sim$75\%) despite [Ne\,\textsc{iii}]/[Ne\,\textsc{ii}] $<1$. The ionization parameter $\gamma$ shows no one-to-one correspondence with the 7.7/11.2~$\mu$m ratio, indicating that PAH ionization is influenced by additional environmental factors. Ionic abundances exhibit both positive and negative correlations with radiation hardness, depending on the species. The limited sample size renders these trends suggestive rather than statistically definitive.

\end{abstract}



%

\section{Introduction} 
\label{section: introduction}

\begin{figure*}
\begin{center}
\resizebox{17.0cm}{17.0cm}{\includegraphics{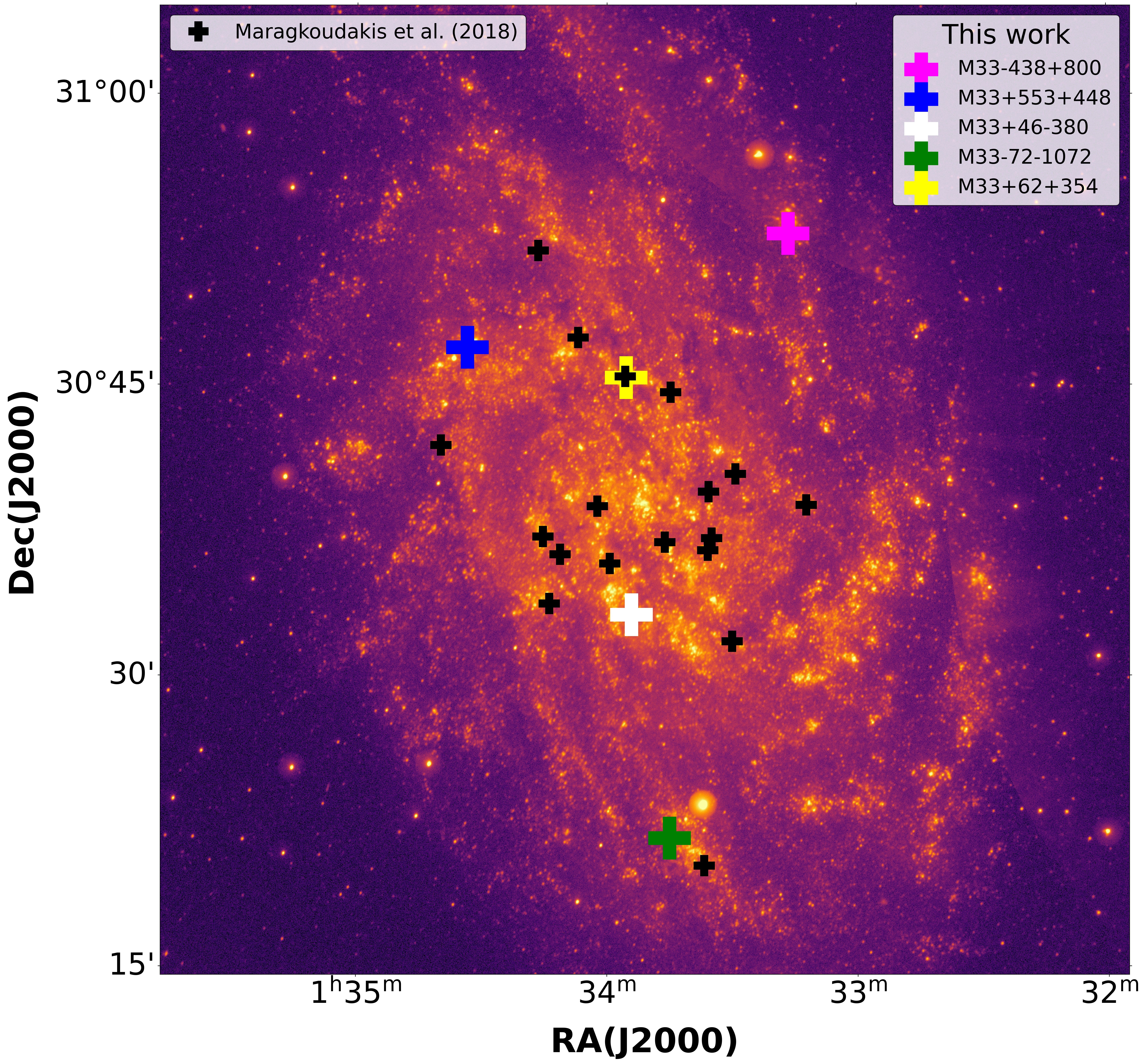}}
\caption{GALEX near-ultraviolet image of the M33 galaxy, observed at an effective wavelength of $\sim$227~nm with an angular resolution of $\sim$5.3~arcsec. The violet, blue, white, green, and yellow plus symbols mark the locations of the five H\,\textsc{ii} regions—M33$-$438+800, M33+553+448, M33+46$-$380, M33$-$72$-$1072, and M33+62+354—analyzed in this work using archival JWST/MIRI observations. The black plus symbols denote the positions of the 18 H\,\textsc{ii} regions previously studied with \textit{Spitzer} by \citet{maragkoudakis2018pahs}.
}\label{Fig: UV image of M33 galaxy}
\end{center}
\end{figure*}

Massive stars ($\gtrsim$8 M$_\odot$; \citealt{bally2008overview}) profoundly reshape their natal molecular environments through their intense ultraviolet (UV) radiation fields. These young OB stars emit copious Lyman-continuum photons, with typical production rates reaching $\sim 10^{50}$ s$^{-1}$ \citep{panagia1973some, stahler2008formation}. The high-energy photons both dissociate H$_2$ and ionize atomic hydrogen, carving out zones of fully ionized gas known as H\,\textsc{ii} regions. Because each ionization event consumes one photon, the size of an H\,\textsc{ii} region around a star with fixed UV output is constrained: for a uniform ambient density, the resulting ionized volume forms the classical Str\"omgren sphere. As massive stars evolve, their powerful radiatively-driven winds can further disrupt and erode these regions by sweeping away the parent cloud material \citep{stahler2008formation}. Consequently, the presence and morphology of an H\,\textsc{ii} region provide a sensitive tracer of very recent or ongoing massive star formation within a molecular cloud.

Within H\,\textsc{ii} regions, recombination between free electrons and ionized hydrogen gives rise to a rich suite of hydrogen recombination lines. The surrounding gas also contains metals inherited from the molecular cloud; these atoms and ions are photoionized by the same UV field and produce numerous ionic emission lines. The relative strengths of these metallic lines encode information about the hardness of the radiation field, the ionization structure, and the local physical conditions of the ionized gas. Importantly, not all H$_2$ molecules are destroyed: a fraction survive in the partially shielded zones and are excited to higher rotational or vibrational levels via UV pumping, generating characteristic ro-vibrational and pure rotational H$_2$ emission features.

More than one hundred molecular species---spanning simple diatomic molecules to complex carbon-chain systems---have been identified in molecular clouds \citep{stahler2008formation}. Among them, polycyclic aromatic hydrocarbons (PAHs) represent a particularly influential family. These large, planar carbonaceous molecules are easily excited by UV photons and radiate prominently in the mid-infrared (MIR), producing a set of well-studied emission bands that serve as powerful diagnostics of the physical conditions in H\,\textsc{ii} regions, photodissociation regions (PDRs), and the diffuse interstellar medium. The strongest PAH features appear at 3.3, 6.2, 7.7, 8.6, 11.3, and 12.7 $\mu$m, with the 3.3 and 11.3 $\mu$m bands tracing predominantly neutral PAHs, while the 6.2, 7.7, and 8.6 $\mu$m bands originate mainly from ionized PAHs \citep{tielens2008interstellar, singh2025jwst, mackie2015characterizing, maltseva2016high}. These bands arise from distinct vibrational modes: the 3.3~$\mu$m band arises from C--H stretching modes, while the 11.3~$\mu$m feature corresponds to C--H out-of-plane bending modes \citep{mackie2015characterizing, maltseva2016high}. The 6.2~$\mu$m feature is primarily associated with C--C stretching modes, whereas the 7.7~$\mu$m feature arises from a combination of C--C stretching and C--H in-plane bending modes \citep{tielens2008interstellar}. The 8.6~$\mu$m band is predominantly attributed to C--H in-plane bending modes. Numerous weaker PAH bands---located at $\sim$3.4, 3.5, 5.25, 5.75, 6.0, 6.9, 7.5, 10.5, 11.0, 13.5, 14.2, 15.8, 16.4, 17.0, 17.4, and 18.9 $\mu$m---also appear in high-quality MIR spectra \citep{maragkoudakis2018pahs, tielens2008interstellar}. Because some bands trace ionized PAHs and others trace neutral PAHs, their relative intensities provide constraints on the mean PAH ionization fraction. Similarly, ratios such as 3.3/11.3 $\mu$m offer insights into the size distribution of PAH molecules \citep{maragkoudakis2020probing}, with shorter-wavelength features preferentially produced by smaller PAHs.

In this work, we investigate a sample of H\,\textsc{ii} regions in the nearby Local Group spiral galaxy M33 (Triangulum Galaxy; \citealt{verley2007star}) using integral-field unit (IFU) spectroscopy obtained with the \emph{James Webb Space Telescope} (JWST) \emph{Mid-Infrared Instrument} (MIRI). The data were retrieved from the \emph{Mikulski Archive for Space Telescopes} (MAST). M33 is located at a distance of $\sim$840 kpc and has an inclination of $i = 56^\circ$ with a position angle of the line of nodes $\theta = 23^\circ$ \citep{maragkoudakis2018pahs}. Its heliocentric velocity is approximately $-179$ km s$^{-1}$ \citep{maryeva2020asymmetrical}. Owing to its proximity, well-resolved structure, and abundance of bright H\,\textsc{ii} regions, M33 has long served as a benchmark system for studies of massive star formation, interstellar medium conditions, and the PAH behavior across diverse galactic environments.

\begin{deluxetable*}{ccccc} \tablecaption{H\,\textsc{ii} regions in M33 analyzed using archival JWST/MIRI Data. \label{table: hii regions}} \tablewidth{0pt} \tablehead{ \colhead{Index} & \colhead{H\,\textsc{ii} Region} & \colhead{RA (J2000)} & \colhead{DEC (J2000)} & \colhead{JWST Proposal ID} } \startdata 1 & M33-438+800 & 01:33:16.500 & +30:52:49.21 & 4297 \\ 2 & M33+553+448 & 01:34:33.500 & +30:46:57.00 & 4297 \\ 3 & M33+46-380 & 01:33:54.100 & +30:33:09.45 & 4297 \\ 4 & M33-72-1072 & 01:33:45.000 & +30:21:38.41 & 4297 \\ 5 & M33+62+354 & 01:33:55.400 & +30:45:23.41 & 4297 \\ \enddata \end{deluxetable*}

Previous investigations have used M33 to explore PAH physics, star-formation tracers, and massive-star feedback. \citet{maragkoudakis2018pahs} analyzed MIR spectra of 18 H\,\textsc{ii} regions in M33, demonstrating how PAH band ratios relate to radiation hardness, metallicity, PAH size distribution, and environment. Their work also showed that commonly used PAH ratios do not always trace PAH size reliably and concluded that extragalactic H\,\textsc{ii} regions are better templates for galaxy-scale PAH emission than many Galactic analogs. \citet{calapa2014heating} examined whether 8 $\mu$m emission reliably traces star formation in M33, finding that it correlates more strongly with cold dust (250 $\mu$m) and evolved stellar emission (3.6 $\mu$m) than with established star-formation tracers such as 24 $\mu$m or H$\alpha$+24 $\mu$m. They argued that PAHs are suppressed in intense star-forming zones and that most 8 $\mu$m emission originates from diffuse ISM heating, making it a poor standalone star formation rate (SFR) indicator. Using Spitzer IRAC/MIPS data, \citet{verley2007star} studied the infrared properties of H\,\textsc{ii} regions, supernova remnants, and planetary nebulae across M33, showing that most 24 $\mu$m sources correspond to H\,\textsc{ii} regions with warm dust and moderate extinction. Their catalog enabled a robust estimate of the galaxy’s IR luminosity function and a consistent star-formation rate of $\sim 0.2~\mathrm{M_\odot~yr^{-1}}$ for the inner disk. Complementing these studies, \citet{massey2006survey} provided deep UBVRI photometry and spectral classifications of massive stars in M31 and M33, improving the census of OB stars and identifying new luminous blue variable (LBV) candidates. \citet{maryeva2020asymmetrical} mapped the ionized gas around the LBV GR 290, revealing the structure and extent of its associated H\,\textsc{ii} region.

More recently, \citet{rogers2026first} presented the first empirical calibration of the mid-infrared abundance diagnostic $\mathrm{Ne}_{23}$ using \textit{JWST}/MIRI observations of H\,\textsc{ii} regions. Their calibration combines MIR fine-structure line measurements of [Ne\,\textsc{ii}]~12.81~$\mu$m and [Ne\,\textsc{iii}]~15.56~$\mu$m, together with the H\,I Hu$\alpha$~12.37~$\mu$m recombination line, with direct-method oxygen abundances, $12+\log(\mathrm{O/H})$, from the CHemical Abundances Of Spirals (CHAOS) project. The study includes \textit{JWST}/MIRI observations of H\,\textsc{ii} regions in M33 obtained under JWST-GO Program 4297, which is also the source of the archival MIRI observations analyzed in this work, as described in Section~\ref{section: data}.

Owing to \textit{JWST}'s substantially larger primary mirror and enhanced spatial and spectral resolution, these data provide significantly higher sensitivity than previous \textit{Spitzer}-based studies, enabling the detection of faint and spatially compact mid-infrared emission features that were previously unresolved or below the \textit{Spitzer} detection limits. The coordinates and JWST proposal IDs of all five regions are listed in Table~\ref{table: hii regions}, and their locations are shown in Figure~\ref{Fig: UV image of M33 galaxy}, overlaid on the GALEX near-ultraviolet (NUV) image of M33 (effective wavelength $\sim$227~nm). Our spectra reveal prominent PAH complexes at 7.7, 8.6, and 11.3~$\mu$m, multiple H$_2$ pure rotational lines, and several ionic fine-structure transitions. Our primary objective is to characterize the physical conditions, ionization structure, and evolutionary states of young star-forming regions through the analysis of mid-infrared diagnostic tracers. In particular, we investigate how variations in the local radiation field influence the excitation and emission properties of PAHs and key ionic species associated with H\,\textsc{ii} regions. By combining these infrared diagnostics with the derived physical parameters, we aim to establish a comprehensive picture of the interplay between stellar radiation and the surrounding interstellar medium. We further derive a range of physical properties of the H\,\textsc{ii} regions, including their characteristic physical conditions and evolutionary indicators, as discussed in detail in Sections~\ref{section: analysis} and~\ref{section: discussion}.

The paper is organized as follows. In Section~\ref{section: data}, we describe the \textit{JWST}/MIRI observations and the data reduction procedures. Section~\ref{section: Spectral Analysis and Measurements} presents the spectral stitching process, continuum modeling, and emission-line and PAH flux measurements. The methods used to derive the physical parameters are described in Section~\ref{section: analysis}. In Section~\ref{section: discussion}, we discuss the astrophysical implications of our results, and finally, Section~\ref{section: summary} summarizes our main findings and conclusions.

\section{JWST MIRI Data} 
\label{section: data}

We obtained the JWST/MIRI observations of M33 from the Mikulski Archive for Space Telescopes (MAST) through the JWST Mission Search interface (Proposal ID 4297; PI: Rogers, Noah Sidney James). The JWST data analyzed in this work can be accessed via \dataset[doi: 10.17909/cv7r-p658]{https://doi.org/10.17909/cv7r-p658}, which provides access to the individual exposures associated with the program for inspection, preview, and download. The archive provides data at several pipeline stages, and for the present work, we used the fully calibrated Level-3 products. These files represent the final stage of the JWST Science Calibration Pipeline and contain exposure-combined, flux-calibrated, and geometrically reconstructed spectral cubes that are ready for scientific analysis (\url{https://outerspace.stsci.edu/spaces/MASTDOCS/pages/113771322/Science+Data+Products}). No additional background subtraction using dedicated off-source observations was performed in this work. The standard fringe-flat correction is included in the Level-3 products used in this work; however, no additional two-dimensional or residual one-dimensional fringe correction was applied. Although residual fringe features are visible in the spectra, they do not affect the identification of the principal spectral features relevant to our analysis, and the main scientific conclusions of this work remain unchanged. The analysis in this work is based on these Level-3 science products, which were processed with version 2.0.1 of the JWST calibration pipeline, using the Calibration Reference Data System (CRDS) version 13.1.13 and the CRDS context \texttt{jwst\_1535.pmap}. The reduction workflow begins with detector-level processing (Stage-1), which includes reference-pixel subtraction, ramp fitting, nonlinearity corrections, and cosmic-ray identification, producing count-rate images. Each exposure then passes through Stage-2 spectroscopic calibration, where flat-fielding, wavelength calibration, photometric calibration, and MIRI-specific corrections such as residual stray-light mitigation and fringe treatment are applied. The final Stage-3 processing combines all calibrated exposures taken in different grating and dichroic settings, maps the IFU slices to sky coordinates, performs outlier rejection, and constructs the three-dimensional spectral cubes for each channel of the Medium-Resolution Spectrometer (MRS). These Level-3 cubes therefore provide continuous spectral coverage across the full MIRI/MRS wavelength range while incorporating the appropriate geometric, photometric, and wavelength calibrations. For details of the data reduction steps, we refer the reader to \cite{singh2025jwst}. \cite{argyriou2023jwst} summarizes the wavelength ranges, resolving powers, and IFU fields of view (FOV) for the four MRS channels.

\section{Spectral Processing and Flux Measurements} 
\label{section: Spectral Analysis and Measurements}

\begin{figure*}
\begin{center}
\resizebox{17.0cm}{6.0cm}{\includegraphics{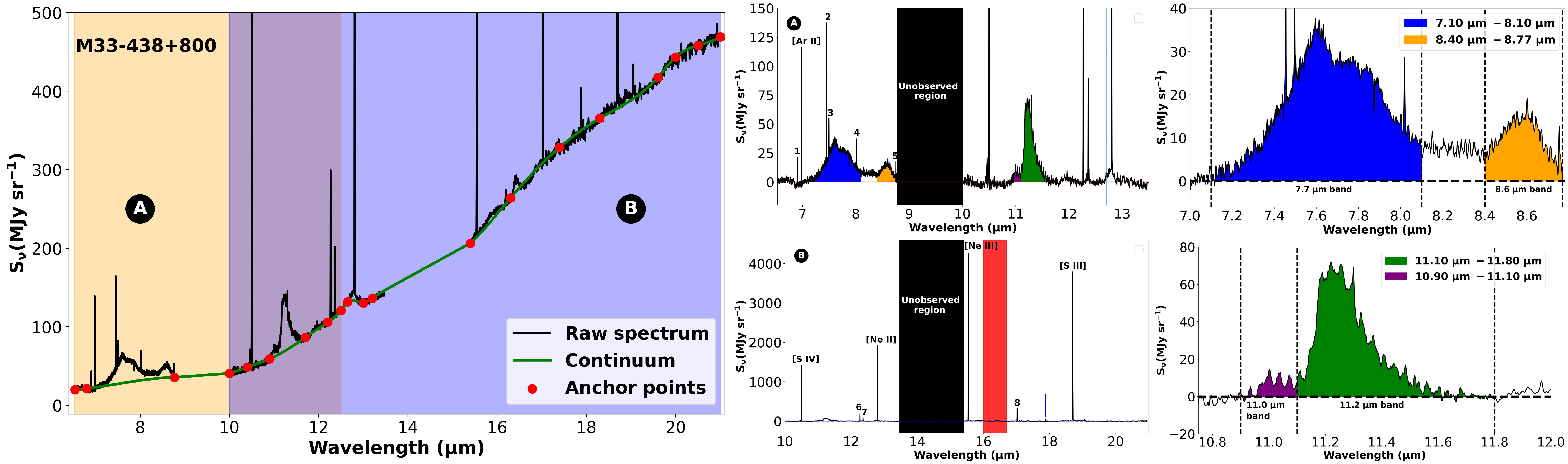}}
\caption{\textbf{Left:} The \textit{JWST}/MIRI MRS raw spectrum of the H\,\textsc{ii} region M33-438+800, covering Channel 1-LONG, 2-SHORT, 2-LONG, 3-SHORT, 3-LONG, and 4-SHORT, spanning a wavelength range of 6.53--20.95~$\mu$m, is shown in black. The red markers indicate the anchor points adopted for continuum determination, while the green curve represents the resulting best-fit continuum model. To highlight the prominent PAH emission features at $\sim$7.7, 8.6, 11.0, and 11.2~$\mu$m, along with key ionic, molecular, and hydrogen recombination lines, the spectrum is partitioned into two segments: Region A (6.53--12.50~$\mu$m) and Region B (10.00--21.00~$\mu$m). \textbf{Middle:} Continuum-subtracted spectra for Regions A (upper panel) and B (lower panel). In the upper panel, the shaded regions in blue, orange, purple, and green delineate the PAH emission bands centered at 7.7, 8.6, 11.0, and 11.2~$\mu$m, respectively. Prominent ionic (metallic) emission lines are labeled above their respective features in both panels. The numerical labels (1--8) denote well-established H$_2$ rotational transitions and hydrogen recombination lines. The red shaded region highlights the weak PAH feature near 16.4~$\mu$m. Of particular interest, the blue vertical tick marks a faint spectral feature that is only rarely reported in $\mathrm{H\,\textsc{ii}}$ regions. To the best of our knowledge, this feature is detected here for the first time in the present source, with a statistically significant signal-to-noise ratio of $\mathrm{S/N}>3$, providing compelling evidence for its presence in the observed spectrum. The black shaded regions indicate wavelength intervals not covered by the \textit{JWST}/MIRI observations. \textbf{Right:} Enlarged views of selected wavelength intervals, 7.00--8.77~$\mu$m (upper panel) and 10.75--12.00~$\mu$m (lower panel), emphasizing the detailed structure of the PAH emission complexes. Dashed vertical lines mark the adopted integration ranges for individual PAH bands: 7.10--8.10~$\mu$m (7.7~$\mu$m complex), 8.40--8.77~$\mu$m (8.6~$\mu$m band), 10.90--11.10~$\mu$m (11.0~$\mu$m band), and 11.10--11.80~$\mu$m (11.2~$\mu$m band).}\label{Fig: M33-438+800 spectrum image}
\end{center}
\end{figure*}

\subsection{MIRI Spectral Stitching and Continuum Modeling}
\label{section: MIRI Spectral Stitching and Continuum Modeling}

The five H\,\textsc{ii} regions analyzed in this study were observed in six JWST/MIRI MRS sub-bands: CH1 Long (Ch1L), CH2 Short (Ch2S), CH2 Long (Ch2L), CH3 Short (Ch3S), CH3 Long (Ch3L), and CH4 Short (Ch4S). Because the FOV of the MIRI MRS increases progressively from Channel~1 to Channel~4, the spatial coverage differs among the individual channels. Consequently, only the region encompassed by the Ch1L observations possesses complete spectral information across all available wavelengths, whereas areas lying outside the Ch1L footprint lack coverage from the shorter-wavelength channels. To ensure uniform spectral sampling and avoid incomplete wavelength coverage, all analyses presented in this work are restricted to the common Ch1L field of view.

Previous studies have generally benefited from complete MIRI MRS wavelength coverage when constructing stitched spectra. For instance, \cite{chown2024pdrs4all} analyzed the Orion Bar using JWST/NIRSpec IFU and MIRI MRS observations, reprojecting all twelve MIRI MRS sub-bands (CH1 Short, CH1 Medium, CH1 Long, CH2 Short, CH2 Medium, CH2 Long, CH3 Short, CH3 Medium, CH3 Long, CH4 Short, CH4 Medium, and CH4 Long) onto a common spatial grid defined by CH1 Short before stitching the spectra using Ch2L as the reference. Likewise, \cite{zhang2025jwst}, in their study of the 30 Doradus star-forming region based on JWST/NIRSpec IFU and MIRI MRS observations, adopted Ch2S as the reference channel for spectral stitching.

Our dataset, however, differs substantially from these studies. It does not include observations in CH1 Medium, CH2 Medium, CH3 Medium, or CH4 Medium, and although CH1 Short, and CH4 Long data are available, they were intentionally excluded because they contain no significant PAH emission relevant to the objectives of this work. Owing to this incomplete wavelength coverage, adopting Ch2S or Ch2L as a global reference is neither necessary nor advantageous.

Instead, all data cubes were first reprojected onto a common spatial grid defined by Ch1L, which represents the region with complete wavelength coverage for all retained observations. Flux normalization was then carried out using a sequential overlap-based scaling procedure that exploits only the wavelength regions shared by adjacent MIRI sub-bands. First, a multiplicative scaling factor was derived by comparing the overlapping spectral interval between Ch1L and Ch2S. This factor was subsequently applied to both Ch2S and Ch2L, thereby placing the entire Channel~2 dataset on the same flux scale as Ch1L. Next, the overlap between the rescaled Ch2L cube and Ch3S was used to determine a second scaling factor, which was then applied to both Ch3S and Ch3L, effectively transferring the established flux calibration to Channel~3. Finally, Ch4S was normalized using the overlap with the rescaled Ch3L cube. By propagating the flux calibration successively through overlapping adjacent channels, this chaining strategy ensures a continuous and internally consistent stitched spectrum without requiring any single intermediate sub-band to serve as a universal reference. Given the absence of several MIRI MRS sub-bands in the JWST/MIRI observations used in this work, this methodology provides a robust and physically motivated approach for preserving relative flux calibration while minimizing the propagation of systematic uncertainties across the full spectral range.

The left panel of Figure~\ref{Fig: M33-438+800 spectrum image} presents the stitched raw spectrum of the H\,\textsc{ii} region M33$-$438$+$800, incorporating all available JWST/MIRI observations for this source. The observed spectrum consists of PAH features together with H$_2$ rotational, HI recombination, and atomic fine-structure emission lines superimposed on a strong dust continuum. To isolate these emission features, the underlying continuum was modeled and removed using a spline interpolation through carefully selected anchor points, following the methodology adopted by \cite{chown2024pdrs4all} and \cite{zhang2025jwst}. The continuum anchor points are indicated by red dots, while the fitted spline continuum is shown by the green curve.

For clarity, the observed spectrum is divided into two wavelength intervals. Region A, spanning $6.53$--$12.50~\mu$m, encompasses the major PAH emission bands at 7.7, 8.6, and 11.2~$\mu$m, together with the weaker 11.0~$\mu$m PAH feature. Region B, extending from $10.00$ to $21.00~\mu$m, covers a rich ensemble of atomic fine-structure lines, molecular emission features, hydrogen recombination lines, and the weaker PAH emission feature centered around 16.4~$\mu$m originated from C-C-C bending modes \citep{tielens2008interstellar}. The continuum-subtracted spectra, along with the PAH features identified in this study, are presented in Figure~\ref{Fig: M33-438+800 spectrum image}; the corresponding spectral features and the wavelength intervals adopted for their integration are detailed in the figure caption.

\begin{figure*}
\begin{center}
\resizebox{17.0cm}{8.0cm}{\includegraphics{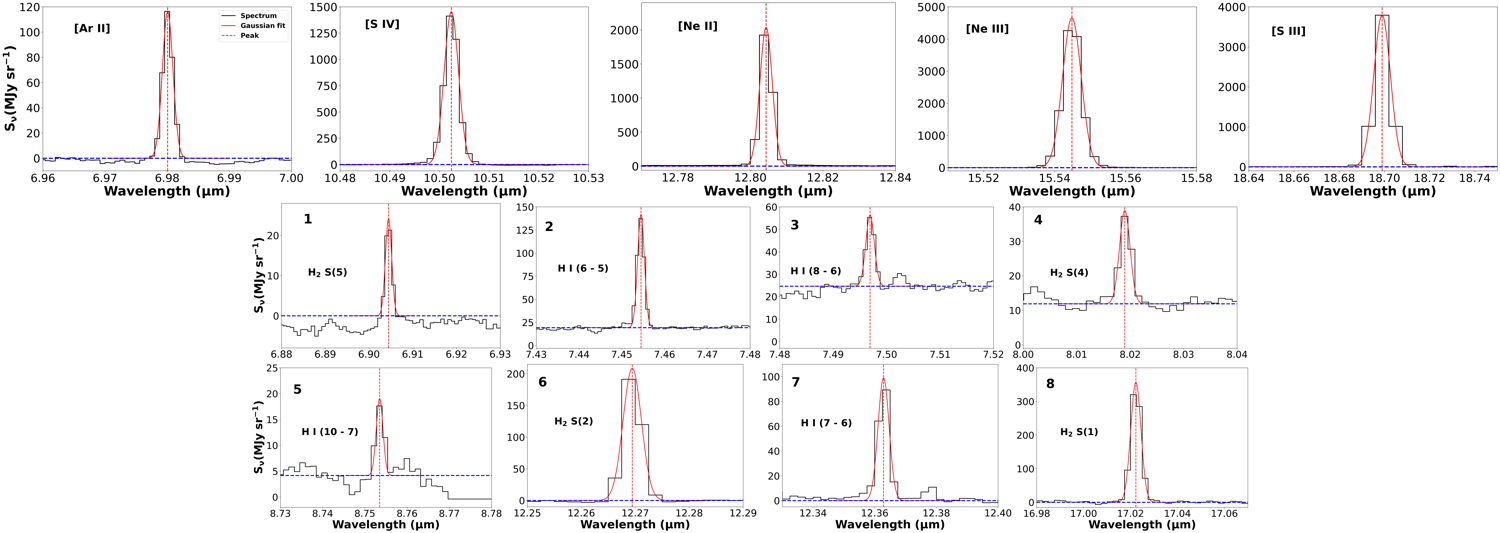}}
\caption{\textbf{Upper:} Spectra of prominent ionic (metallic) emission lines detected toward the H\,\textsc{ii} region M33-438+800. \textbf{Lower:} Spectra of hydrogen molecular rotational transitions and hydrogen recombination lines, labeled 1--8 as defined in Figure~\ref{Fig: M33-438+800 spectrum image}. In both panels, the observed spectra are shown in black, while the red curves represent the corresponding Gaussian fits to the line profiles. The red dashed vertical lines indicate the centroid (peak) wavelengths of the fitted features, and the blue dashed horizontal lines denote the baseline levels adopted for the Gaussian fitting.}\label{Fig: M33-438+800 metallic spectrum}
\end{center}
\end{figure*}

Figure~\ref{Fig: M33-438+800 metallic spectrum} shows the continuum-subtracted profiles of the atomic, H$_2$ rotational, and HI recombination emission lines for M33$-$438$+$800. The best-fitting Gaussian models are overplotted in red, and the local baselines are indicated by blue dashed lines. Equivalent spectral analyses for the remaining four H\,\textsc{ii} regions are presented in Appendix~\ref{section: PAH Spectra of the Remaining Four H II Regions} and~\ref{section: Metallicity, H2, and HI Spectra of the Remaining Four H II Regions}.

We identify two narrow emission features at approximately 7.879 and 17.885~$\mu$m. The feature at 17.885~$\mu$m is consistently detected across all five H\,\textsc{ii} regions, whereas the 7.879~$\mu$m feature is detected only in the second, fourth, and fifth regions. The former is considered a robust detection based on its signal-to-noise ratio of $\mathrm{S/N}>3$. For features superposed on the broad PAH emission bands, however, the underlying continuum and band structure complicate a reliable local noise estimate. In such cases, we conservatively regard a feature as detectable when it is resolved by at least five independent spectral points and exhibits a distinct spectral profile above the underlying PAH emission.

The feature at 17.885~$\mu$m is identified as the forbidden [P\,\textsc{iii}] fine-structure transition, consistent with the identification reported by \citet{rogers2026first}. Although this line has been reported in a limited number of astrophysical environments, primarily planetary nebulae \citep{pottasch2008abundances}, it has rarely been identified in H\,\textsc{ii} regions, owing to the intrinsically low cosmic abundance of phosphorus and the resulting weakness of the line. The detection of this feature across all five H\,\textsc{ii} regions in M33 demonstrates the capability of \textit{JWST}/MIRI to detect this weak ionic transition in extragalactic star-forming environments.

In contrast, the emission feature at 7.879~$\mu$m could not be matched to any known molecular, atomic or ionic transition listed in the \textit{Infrared Space Observatory} (ISO) infrared spectral line catalog provided by the Max Planck Institute for Extraterrestrial Physics (MPE), which served as the primary reference for identifying the remaining emission lines detected in our spectra. The absence of a corresponding transition in the available line lists suggests that this feature may arise from a previously unreported or currently unidentified emission component. Additional observational and theoretical investigations will be required to establish its physical origin.

\subsection{Flux Measurement of PAH Bands and Narrow Emission Lines}
\label{section: Flux Measurement of PAH Bands and Narrow Emission Lines}

To determine the integrated fluxes of PAH emission features, we adopted the PAH band integration ranges defined by \cite{zhang2025jwst}. Accordingly, the 7.7~$\mu$m PAH complex was integrated over 7.10--8.10~$\mu$m, the 8.6~$\mu$m feature over 8.40--8.77~$\mu$m, the 11.0~$\mu$m feature over 10.90--11.10~$\mu$m, and the 11.2~$\mu$m feature over 11.10--11.80~$\mu$m. Similar to \cite{zhang2025jwst}, we do not decompose the blended PAH emission complexes in the 7--9~$\mu$m and 10--15~$\mu$m wavelength ranges into individual subcomponents. Instead, after removing the superposed narrow emission features, we determine the integrated fluxes of the broad PAH complexes directly over the adopted wavelength intervals.

We note that \cite{zhang2025jwst} adopted a wider integration interval of 8.40--9.05~$\mu$m for the 8.6~$\mu$m PAH band. However, the JWST/MIRI observations used in this work do not include the Ch2 Medium sub-band and are limited to Ch2S, restricting the available wavelength coverage to 8.77~$\mu$m. Consequently, the adopted integration range for the 8.6~$\mu$m feature is necessarily truncated, and the derived flux may underestimate the total band strength by excluding emission beyond 8.77~$\mu$m. This limitation arises solely from the incomplete observational wavelength coverage and should therefore be considered when comparing our measured 8.6~$\mu$m PAH fluxes with studies based on complete MIRI MRS datasets.

Prior to measuring the PAH fluxes, the narrow atomic, molecular, and recombination emission lines superposed on the broad PAH features were identified and removed from the continuum-subtracted spectra to isolate the intrinsic PAH emission. The integrated PAH fluxes were then obtained by numerically integrating the residual spectra over the adopted wavelength intervals. The resulting integrated fluxes for the three major PAH bands at 7.7, 8.6, and 11.2~$\mu$m, together with the weaker 11.0~$\mu$m PAH feature, are presented in Table~\ref{table: PAH emission}. The 16.4~$\mu$m PAH feature is relatively weak in our sample, as revealed by the archival \textit{JWST}/MIRI observations. The feature is detected in four of the five H\,\textsc{ii} regions, with integrated fluxes of approximately $1.1 \times 10^{-17} \pm 7.4 \times 10^{-19}$~W~m$^{-2}$, $1.9 \times 10^{-17} \pm 8.6 \times 10^{-19}$~W~m$^{-2}$, $1.5 \times 10^{-18} \pm 8.2 \times 10^{-20}$~W~m$^{-2}$, and $4.7 \times 10^{-18} \pm 3.4 \times 10^{-19}$~W~m$^{-2}$ for M33-438+800, M33+553+448, M33+46-380, and M33+62+354, respectively. Among the PAH bands considered in this study, the 16.4~$\mu$m feature contributes only $\sim$1\% of the total PAH emission in M33-438+800 and M33+553+448, and less than 1\% in M33+46-380 and M33+62+354. Its integrated flux is also approximately a factor of two lower than that of the weaker 11.0~$\mu$m PAH band, further indicating that the 16.4~$\mu$m feature represents the lowest contribution to the overall PAH emission in these regions.

Furthermore, the physical properties and PAH diagnostic quantities derived in this work do not explicitly depend on the integrated flux of the 16.4~$\mu$m feature. Given its relatively small contribution to the overall PAH emission and its limited relevance to the quantitative diagnostics considered here, we therefore do not include the 16.4~$\mu$m feature in the subsequent quantitative analysis.

For completeness, we also examined the wavelength interval used for the 16.4~$\mu$m feature in the literature. \citet{boersma201015} considered the 16.0--16.6~$\mu$m interval in their analysis of the 16.4~$\mu$m PAH feature using lower-resolution \textit{Spitzer}/IRS observations and PAH model spectra. However, when applied to our higher-resolution \textit{JWST}/MIRI spectra, we find that the 16.4~$\mu$m emission is predominantly confined to a narrower interval of approximately 16.30--16.52~$\mu$m. Outside this wavelength range, the spectra either become dominated by noise or show no appreciable emission associated with the 16.4~$\mu$m feature. We therefore adopted 16.30--16.52~$\mu$m as the integration interval for the 16.4~$\mu$m feature in our measurements. This choice allows us to isolate the observed feature more appropriately in the higher spectral-resolution \textit{JWST}/MIRI data while avoiding regions that do not show significant associated emission.

To the best of our knowledge, there is currently no widely established integration prescription specifically defining the wavelength limits of the 16.4~$\mu$m PAH feature for \textit{JWST}/MIRI observations. This differs from the treatment of the shorter-wavelength PAH features, for which studies such as \citet{zhang2025jwst} provide explicit integration intervals over the 3--15~$\mu$m wavelength range. A systematic investigation of the 15--20~$\mu$m PAH features with \textit{JWST}/MIRI, including a detailed definition of the appropriate integration ranges, would therefore be valuable for future studies; such an investigation is beyond the scope of the present work.

The fluxes of the narrow atomic fine-structure, H$_2$ rotational, and HI recombination lines were measured using a Gaussian-fitting procedure. For each emission feature, a Gaussian profile was fitted to the continuum-subtracted spectrum to determine the centroid wavelength ($\lambda_{\rm mean}$) and the standard deviation ($\sigma$). The full width at half maximum (FWHM) was computed as $\mathrm{FWHM}=2.35\sigma$, and the integrated line flux was subsequently evaluated over the interval $\lambda_{\rm mean}\pm\mathrm{FWHM}$. Adopting an integration window based on the fitted line width provides a consistent and robust estimate of the total line flux while minimizing contamination from the surrounding continuum and adjacent spectral features. The measured integrated fluxes of the atomic fine-structure lines, H$_2$ rotational transitions, and HI recombination lines are listed in Tables~\ref{table: metallic emission}--\ref{table: hydrogen recombination emission}.

\setlength{\tabcolsep}{3pt} 

\begin{table*}[ht]
\scriptsize
\caption{Integrated fluxes of the diagnostic PAH bands obtained for the five H\,\textsc{ii} regions in M33.}
\label{table: PAH emission}

\raggedright

\begin{tabular}{ccccccc}
\hline
Index & \multicolumn{4}{c}{Integrated Flux (W\,m$^{-2}$)} & F(7.7)/F(11.2) & F(8.6)/F(11.2)\\
      & 7.7 $\mu$m & 8.6 $\mu$m & 11.0 $\mu$m & 11.2 $\mu$m &  & \\
\hline

1 & $5.11E{-}16 \pm 1.42E{-}18$ & $8.77E{-}17 \pm 7.88E{-}19$ & $1.60E{-}17 \pm 5.88E{-}18$ & $2.01E{-}16 \pm 1.68E{-}17$ & $2.54 \pm 0.21$ & $0.44 \pm 0.04$\\

2 & $1.02E{-}15 \pm 1.76E{-}18$ & $1.68E{-}16 \pm 9.41E{-}19$ & $3.82E{-}17 \pm 5.51E{-}18$ & $4.36E{-}16 \pm 1.66E{-}17$ & $2.34 \pm 0.09$ & $0.39 \pm 0.02$\\

3 & $1.97E{-}16 \pm 8.30E{-}19$ & $4.31E{-}17 \pm 4.70E{-}19$ & $6.14E{-}18 \pm 1.77E{-}18$ & $8.54E{-}17 \pm 4.98E{-}18$ & $2.31 \pm 0.14$ & $0.51 \pm 0.03$\\

4 & $8.27E{-}17 \pm 9.18E{-}19$ & $1.97E{-}17 \pm 4.86E{-}19$ & $3.96E{-}18 \pm 2.25E{-}18$ & $4.12E{-}17 \pm 7.14E{-}18$ & $2.01 \pm 0.35$ & $0.46 \pm 0.08$\\

5 & $4.03E{-}16 \pm 1.52E{-}18$ & $7.87E{-}17 \pm 8.12E{-}19$ & $1.10E{-}17 \pm 2.81E{-}18$ & $1.59E{-}16 \pm 7.91E{-}18$ & $2.54 \pm 0.13$ & $0.50 \pm 0.03$\\

\hline
\end{tabular}

\end{table*}

\setlength{\tabcolsep}{3pt} 

\begin{table*}[ht]
\scriptsize  
\caption{Integrated fluxes of ionic metal emission lines detected within the five H\,\textsc{ii} regions in M33.}
\label{table: metallic emission}

\raggedright  

\begin{tabular}{ccccccc}
\hline
Index & \multicolumn{5}{c}{Integrated Flux (W\,m$^{-2}$)} \\
      & [Ar II] & [Ne II] & [Ne III] & [S III] & [S IV] \\
\hline

1 & $8.65E{-}18 \pm 1.18E{-}19$ & $8.47E{-}17 \pm 1.93E{-}19$ & $2.12E{-}16 \pm 7.12E{-}19$ & $1.41E{-}16 \pm 2.90E{-}19$ & $8.15E{-}17 \pm 4.93E{-}19$ \\

2 & $3.80E{-}17 \pm 2.39E{-}19$ & $2.31E{-}16 \pm 4.22E{-}19$ & $2.13E{-}16 \pm 9.27E{-}19$ & $3.36E{-}16 \pm 9.61E{-}19$ & $6.05E{-}17 \pm 3.72E{-}19$ \\

3 & $1.24E{-}17 \pm 1.33E{-}19$ & $6.21E{-}17 \pm 9.51E{-}20$ & $1.43E{-}17 \pm 5.26E{-}20$ & $4.86E{-}17 \pm 7.40E{-}20$ & $3.37E{-}18 \pm 4.65E{-}20$ \\

4 & $2.00E{-}18 \pm 5.65E{-}20$ & $1.22E{-}17 \pm 4.97E{-}20$ & $2.00E{-}17 \pm 8.46E{-}20$ & $2.26E{-}17 \pm 9.91E{-}20$ & $6.16E{-}18 \pm 5.71E{-}20$ \\

5 & $3.48E{-}17 \pm 2.08E{-}19$ & $1.49E{-}16 \pm 3.25E{-}19$ & $3.18E{-}17 \pm 1.47E{-}19$ & $1.19E{-}16 \pm 2.40E{-}19$ & $8.26E{-}18 \pm 6.01E{-}20$ \\

\hline
\end{tabular}
\end{table*}

\setlength{\tabcolsep}{3pt} 

\begin{table*}[ht]
\caption{Integrated fluxes of H$_2$ pure rotational lines for the five H\,\textsc{ii} regions in M33.}
\label{table: hydrogen emission}

\raggedright

\begin{tabular}{cccccc}
\hline
Index & \multicolumn{4}{c}{Integrated Flux (W\,m$^{-2}$)} & \\
      & S(1) & S(2) & S(4) & S(5) \\
\hline

1 & $1.15E{-}17 \pm 6.58E{-}20$ & $9.46E{-}18 \pm 5.40E{-}20$ & $1.86E{-}18 \pm 9.10E{-}20$ & $1.53E{-}18 \pm 9.82E{-}20$ \\

2 & $1.01E{-}17 \pm 1.01E{-}19$ & $7.48E{-}18 \pm 3.98E{-}20$ & $1.42E{-}18 \pm 1.43E{-}19$ & $2.12E{-}18 \pm 9.02E{-}20$ \\

3 & $2.53E{-}18 \pm 3.74E{-}20$ & $1.70E{-}18 \pm 1.23E{-}20$ & $2.96E{-}19 \pm 2.64E{-}20$ & $5.40E{-}19 \pm 2.62E{-}20$ \\

4 & $1.78E{-}18 \pm 3.08E{-}20$ & $1.16E{-}18 \pm 2.52E{-}20$ & $2.98E{-}19 \pm 4.28E{-}20$ & $3.57E{-}19 \pm 5.13E{-}20$ \\

5 & $5.60E{-}18 \pm 6.79E{-}20$ & $4.26E{-}18 \pm 2.41E{-}20$ & $6.51E{-}19 \pm 5.62E{-}20$ & $1.18E{-}18 \pm 5.96E{-}20$ \\

\hline
\end{tabular}


\end{table*}

\setlength{\tabcolsep}{3pt} 

\begin{table*}[ht]
\caption{Integrated H I recombination line fluxes detected for all five  H\,\textsc{ii} regions in M33 galaxy.}
\label{table: hydrogen recombination emission}

\raggedright  

\begin{tabular}{ccccc}
\hline
Index & \multicolumn{4}{c}{Integrated Flux (W\,m$^{-2}$)} \\
     & (6--5) & (10--7) & (7--6) & (8--6) \\
\hline

1 & $7.53E{-}18 \pm 1.96E{-}19$ & $5.81E{-}19 \pm 6.84E{-}20$ & $4.67E{-}18 \pm 1.24E{-}19$ & $1.89E{-}18 \pm 1.22E{-}19$ \\

2 & $1.68E{-}17 \pm 6.25E{-}19$ & $1.17E{-}18 \pm 4.87E{-}20$ & $5.84E{-}18 \pm 1.38E{-}19$ & $4.16E{-}18 \pm 3.07E{-}19$ \\

3 & $2.34E{-}18 \pm 1.38E{-}19$ & $1.71E{-}19 \pm 3.08E{-}20$ & $1.15E{-}18 \pm 3.80E{-}20$ & $5.55E{-}19 \pm 5.14E{-}20$ \\

4 & $1.34E{-}18 \pm 5.76E{-}20$ & $1.81E{-}19 \pm 1.81E{-}19$ & $4.27E{-}19 \pm 2.06E{-}20$ & $3.62E{-}19 \pm 5.52E{-}20$ \\

5 & $5.56E{-}18 \pm 1.26E{-}19$ & $3.80E{-}19 \pm 3.92E{-}20$ & $2.52E{-}18 \pm 5.38E{-}20$ & $1.20E{-}18 \pm 1.08E{-}19$ \\

\hline
\end{tabular}


\end{table*}

\section{Analysis} 
\label{section: analysis}

\subsection{Rotation Diagram Analysis of the M33 H\,\textsc{ii} Regions}
\label{section: Rotation Diagram Analysis of the M33 H II Regions}

The pure rotational emission lines of molecular hydrogen provide a powerful and direct diagnostic of warm molecular gas in star-forming environments. As discussed by \citet{franceschi2024minds}, these transitions can be reasonably treated under the assumption of local thermodynamic equilibrium (LTE). This approximation may well justified, as the critical densities ($n_{crit}$) of the H$_2$ 0--0 S($J$) rotational transitions are typically one or more orders of magnitude lower than those of the commonly observed rotational transitions of carbon monoxide (CO), the second most abundant molecular species in the interstellar medium, with characteristic values of $n_{\rm crit} \sim 10^{3}\,\mathrm{cm^{-3}}$ \citep{stahler2008formation}. In addition, the Einstein A coefficients ($A_{J}$) of the H$_{2}$ rotational transitions are considerably smaller than those of CO, enabling the H$_{2}$ level populations to thermalize under comparatively low-density conditions.

Under the LTE approximation and assuming a Boltzmann distribution of level populations, the optically thin H$_2$ rotational line emission satisfies the standard rotation-diagram expression \citep{franceschi2024minds, goldsmith1999population}:
\begin{equation}
    \ln\left(\frac{4\pi F_{J}}{hc\nu_{J} g_{J} A_{J}}\right)
    = -\frac{E_{J}}{k_{B} T_{\rm rot}}
    - \ln\left(\frac{Q_{\rm rot}}{N_{\rm tot}\Omega}\right),
    \label{equation: rotation diagram}
\end{equation}
where $F_J$ is the observed line flux, $\nu_J$ is the transition frequency, $g_J = 2J+1$ is the statistical weight, and $E_J$ is the upper-level energy. The quantities $T_{\rm rot}$, $Q_{\rm rot}$, $N_{\rm tot}$, and $\Omega$ represent the rotational temperature, rotational partition function, total column density, and solid angle of the emitting region, respectively. The assumed ortho-to-para ratio is 3 \citep{franceschi2024minds}. The adopted values of $\nu_J$, $A_J$, and $E_J$ are listed in Table~\ref{table: H2 properties}, while the measured line fluxes are given in Table~\ref{table: hydrogen emission}.

To construct the rotation diagrams, we define  
\begin{equation}
y \equiv \ln\!\left(\frac{N_J}{g_J}\right), \qquad
N_J = \frac{4\pi F_J}{h c \nu_J A_J}, \qquad
x \equiv \frac{E_J}{k_B},
\end{equation}

\begin{figure*}
\begin{center}
\resizebox{16.0cm}{12.0cm}{\includegraphics{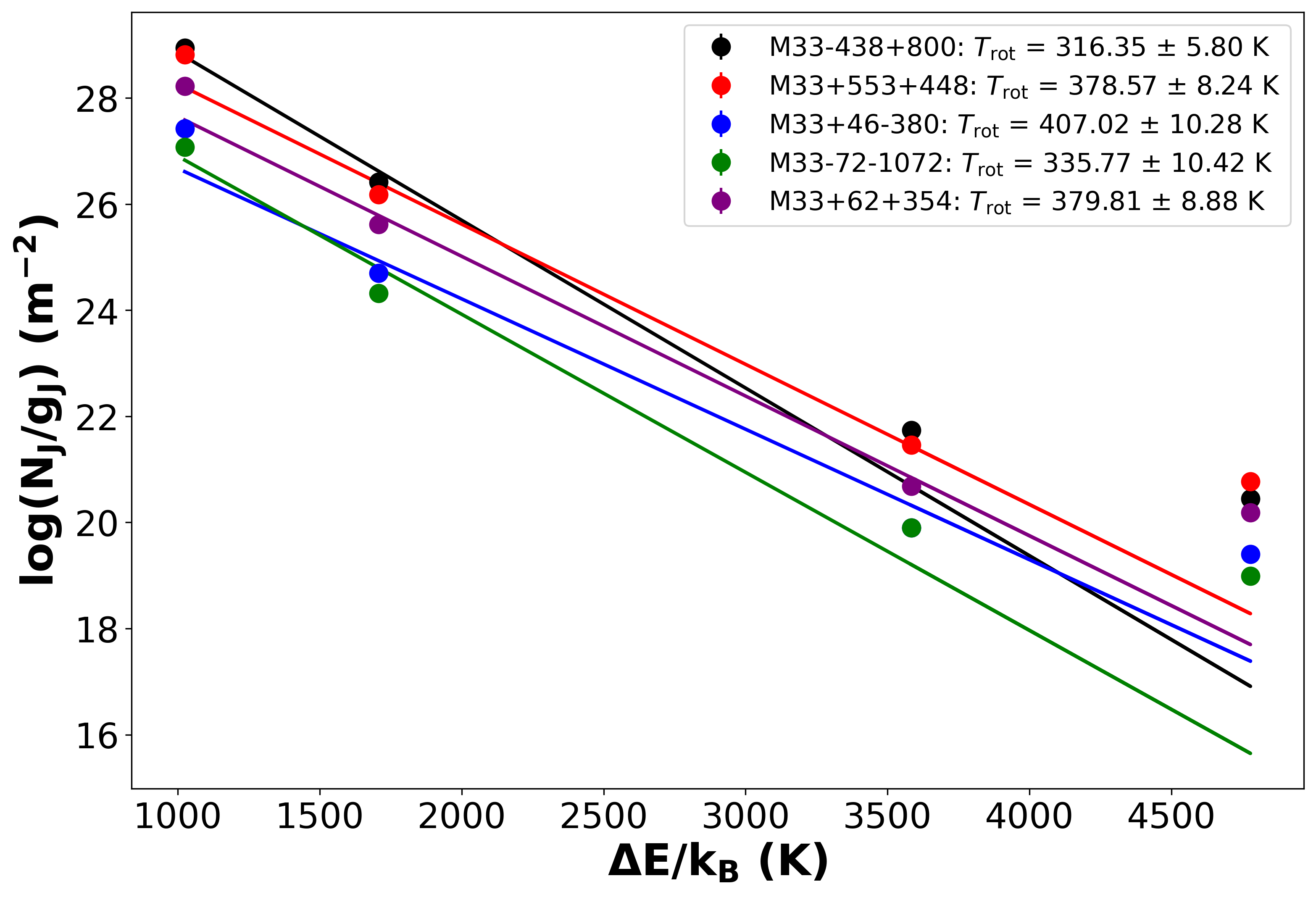}}
\caption{Rotational diagrams for the five H\,\textsc{ii} regions in the M33 galaxy, illustrating the population distribution across molecular rotational energy levels along with their corresponding weighted linear fits.}
\label{Fig: rotation diagram image}
\end{center}
\end{figure*}

Equation~\ref{equation: rotation diagram} therefore describes a straight line in the $(x,y)$ plane with slope $-1/T_{\rm rot}$, and intercept $ln\left(N_{tot}\Omega/Q_{rot}\right)$. For each H\,\textsc{ii} region, we perform a weighted least-squares linear regression, incorporating line-flux uncertainties to determine the optimal slope and intercept. The resulting fit parameters are summarized in Table~\ref{table: H2 line fit}. The corresponding rotation diagrams are shown in Figure~\ref{Fig: rotation diagram image}.

The rotational temperature is derived from the slope, a, of the linear fit according to, $T_{rot} = -\frac{1}{a}$.

The total warm H$_2$ gas column density for each H\,II region was derived from the intercept of the rotation diagram using 
$N_{\rm tot} = \frac{Q_{\rm rot}\,e^{b}}{\Omega}$, 
where $b$ is the intercept of the best-fitting linear relation. The warm H$_2$ gas mass was subsequently estimated using

\begin{equation}
    M = \mu_{\rm H_2}\, m_{\rm H}\, N_{\rm tot}\, A,
\end{equation}
where $\mu_{\rm H_2}=2.8$ is the adopted mean molecular weight \citep{kauffmann2008mambo}, $m_{\rm H}$ is the mass of a hydrogen atom, and $A$ is the projected physical area covered by the JWST/MIRI MRS Ch1L observations for each H\,\textsc{ii} region. The effective radius of the analyzed region is defined as $r_{\rm eff} = \sqrt{\frac{A}{\pi}}$, yielding $r_{\rm eff} \approx 11.3$ pc for the distance of 840 kpc to the galaxy \citep{maragkoudakis2018pahs}. This effective radius is significantly smaller than that adopted for the H\,\textsc{ii} regions within the same galaxy in the \citet{maragkoudakis2018pahs} study based on \textit{Spitzer} observations, owing to the superior spatial resolution and more localized spatial coverage provided by JWST. Consequently, the derived warm H$_2$ column densities and gas masses presented in this work correspond to the compact regions sampled by the JWST observations and are not directly comparable to values obtained from the larger apertures employed in the lower-resolution \textit{Spitzer} data. The derived physical properties of the warm molecular gas, including the rotational temperature, total H$_2$ column density , and warm H$_2$ mass, are listed in Table~\ref{table: HII properties}.

Although the spectra are extracted
from H\,\textsc{ii} regions, the detected H$_2$ pure rotational emission
originates in the surrounding photon-dominated regions (PDR) at the
ionized--molecular interface, where far-ultraviolet (FUV) radiation from
massive stars heat the molecular gas. The H$_2$ rotation diagrams
exhibit approximately linear trends, with RMS scatters ranging from
$\sigma_{Y/X} \simeq 1.1$ to $1.8$. The scatter is defined as the root-mean-
square dispersion of the data points about the best-fit linear relation
in the rotation diagram, $\sigma_{Y/X} =
\left[\langle\left( y_i - (a x_i + b) \right)^2\rangle\right]^{1/2}$ \citep{maragkoudakis2018pahs}. Regions with lower scatter may broadly consistent with a dominant single-temperature LTE component, whereas larger dispersions likely reflect temperature gradients, multiple excitation components, or unresolved substructure within the PDR gas. If the gas is subthermally excited, departures from LTE can preferentially depopulate the higher-$J$ rotational levels, leading to deviations from a linear rotational diagram and, in many cases, yielding a rotational temperature lower than the true kinetic temperature. Consequently, under such conditions, the LTE-derived rotational temperatures may underestimate the kinetic temperature of the warm molecular gas and should therefore be regarded as lower limits rather than direct measurements of $T_{\rm kin}$. Because the total column density depends on both the fitted intercept and the temperature-dependent rotational partition function, any departures from LTE could introduce systematic uncertainties into the derived warm H$_2$ column densities and masses.

A rigorous assessment of these effects would require dedicated non-LTE excitation modelling together with additional observational constraints, such as independent measurements of the gas density and kinetic temperature or observations of complementary molecular tracers (e.g., CO, $^{13}$CO, or C$^{18}$O). In principle, non-LTE radiative transfer codes such as RADEX can be used to investigate excitation conditions for many molecular species when the necessary collisional rate coefficients and molecular data are available. However, the pure rotational H$_2$ transitions analyzed in this work are not routinely treated within the standard RADEX/LAMDA framework \citep{van2007computer, schoier2005atomic}, and the ancillary observational constraints required for a robust non-LTE analysis are not available for the present sample. Consequently, the physical quantities presented here should be regarded as LTE-based estimates that may provide a self-consistent basis for comparing the observed H\,II regions, while recognizing that systematic uncertainties associated with non-LTE excitation and unresolved multi-temperature components may remain.

\begin{table}[ht]
\centering
\caption{
Physical properties of the observed H$_2$ pure rotational transitions. 
Wavelengths and frequencies correspond to the 0--0 S($J$) lines. 
Einstein $A$ coefficients are adopted from \citet{wolniewicz1998quadrupole}. 
Energies refer to the upper-state rotational levels.
\label{table: H2 properties}}
\begin{tabular}{ccccc}
\hline
Transition & $\lambda$ ($\mu$m) & $\nu$ (THz) & $A$ (s$^{-1}$) & $E$ (eV) \\
\hline
S(1) & 17.035 & 17.611 & 4.76E$-$10 & 0.088 \\
S(2) & 12.279 & 24.433 & 2.75E$-$9  & 0.147 \\
S(4) & 8.025  & 37.383 & 2.64E$-$8  & 0.309 \\
S(5) & 6.910  & 43.418 & 5.88E$-$8  & 0.412 \\
\hline
\end{tabular}
\end{table}

\begin{table}[ht]
\centering
\caption{Weighted linear-fit parameters for the H$_2$ rotation diagrams of the five M33 H\,\textsc{ii} regions.}
\label{table: H2 line fit}

\begin{tabular}{ccc}
\hline
Index & Slope & Intercept\\
\hline
1  & $-3.16E-3 \pm 5.80E-5$ & $32.02 \pm 0.11$\\
2  & $-2.64E-3 \pm 5.80E-5$ & $30.91 \pm 0.12$\\
3  & $-2.46E-3 \pm 6.20E-5$ & $29.13 \pm 0.13$ \\
4  & $-2.98E-3 \pm 9.20E-5$ & $29.88 \pm 0.18$\\
5  & $-2.63E-3 \pm 6.20E-5$ & $30.28 \pm 0.13$\\
\hline
\end{tabular}

\end{table}

\begin{table*}
\scriptsize
\centering
\caption{Derived physical parameters of the observed H\,II regions.}
\label{table: HII properties}
\renewcommand{\arraystretch}{1.3}

\begin{tabular}{ccccccccccccc}
\hline
Index &
$T_{\rm rot}$ &
$N_{\rm tot}(\mathrm{H}_2)$ &
Mass &
$\langle Q(\mathrm{H}^{0}) \rangle$ &
$II$ & Ne$^{+2}$/Ne$^{+}$ & S$^{+3}$/S$^{+2}$ & Ne$^{+2}$/S$^{+2}$ & Ne$^{+2}$/S$^{+3}$ & Ne$^{+}$/S$^{+2}$ & Ne$^{+}$/S$^{+3}$ & $Z$ \\
&
(K) &
($10^{19}$ cm$^{-2}$) &
($10^{2}\,M_\odot$) &
($10^{50}$ s$^{-1}$) & & & & & & & &
($Z_\odot$)
\\
\hline

1 & $316.35 \pm 5.80$ & $11.45 \pm 1.22$ & $10.33 \pm 1.10$ & 3.43 & $0.98 \pm 0.03$ & 1.21 & 0.16 & 9.88 & 61.20 & 8.16 & 50.57 & 0.32 \\

2 & $378.57 \pm 8.24$ & $4.44 \pm 0.52$ & $4.00 \pm 0.47$ & 5.90 & $-0.18 \pm 0.05$ & 0.44 & 0.05 & 4.15 & 82.69 & 9.36 & 186.32 & 0.54 \\

3 & $407.02 \pm 10.28$ & $0.80 \pm 0.11$ & $0.72 \pm 0.10$ & 0.94 & $-1.38 \pm 0.07$ & 0.11 & 0.02 & 1.93 & 100.05 & 17.36 & 898.51 & 0.39 \\

4 & $335.77 \pm 10.42$ & $1.43 \pm 0.26$ & $1.29 \pm 0.23$ & 0.56 & $0.33 \pm 0.04$ & 0.79 & 0.08 & 5.78 & 75.94 & 7.31 & 96.05 & 0.51\\

5 & $379.81 \pm 8.88$ & $2.39 \pm 0.30$ & $2.15 \pm 0.27$ & 2.10 & $-1.41 \pm 0.07$ & 0.10 & 0.02 & 1.75 & 90.41 & 16.95 & 876.24 & 0.43 \\

\hline
\end{tabular}
\end{table*}

\subsection{Ionizing Photon Output of the M33 H\,\textsc{ii} Regions}
\label{section: ionizing Photon Output of the HII Regions}

Within H\,\textsc{ii} regions, hydrogen recombination lines provide a powerful diagnostic for estimating the ionizing photon flux produced by massive stars. Under the widely adopted Case~B approximation---originally formulated by \citet{storey1987recombination} and \citet{osterbrock1989astrophysics}---Lyman-series photons are assumed to be optically thick, becoming locally absorbed and reprocessed into lower-energy transitions or two-photon continuum emission. All higher-order (Balmer and above) recombination photons, however, are considered optically thin and freely escape the nebula. In this regime, every ionizing photon emitted by a massive star ultimately results in a recombination to an excited level $n \geq 2$, establishing a direct equality between the total hydrogen-ionizing photon rate, $Q({\rm H}^{0})$, and the total recombination rate to these levels \citep{stahler2008formation}.

For a specific hydrogen recombination line corresponding to the transition $n_{u} \rightarrow n_{l}$, the number of emitted photons per second is
\begin{equation}
N_{\rm rec}(n_{u} \rightarrow n_{l}) = \alpha_{\rm eff}(n_{u} \rightarrow n_{l})\, n_{e}\, n_{p}\, V,
\label{equation: n to n-1}
\end{equation}
where $\alpha_{\rm eff}$ is the effective recombination coefficient for that transition, $n_{u}$, and $n_{l}$ are the principal quantum numbers of upper and lower energy levels respectively, $n_{e}$, $n_{p}$, and $V$ denote the electron density, proton density, and emitting volume, respectively. The total number of recombinations to all excited states ($n \geq 2$) is given by
\begin{equation}
N_{\rm rec}(n \geq 2) = \alpha_{B}\, n_{e}\, n_{p}\, V,
\label{equation: all transitions}
\end{equation}
where $\alpha_{B}$ is the Case~B recombination coefficient. The observed flux of the $n_{u} \rightarrow n_{l}$ transition, $F_{\rm obs}$, is related to the emitted photon rate through
\begin{equation}
F_{\rm obs} = \frac{N_{\rm rec}(n_{u} \rightarrow n_{l})\, h\nu}{4\pi D^{2}},
\label{equation: observed flux}
\end{equation}
with $D$ the distance to the emitting region. Taking the ratio of the above expressions~\ref{equation: n to n-1} and~\ref{equation: all transitions} and using $Q({\rm H}^{0}) = N_{\rm rec}(n \geq 2)$ yields
\begin{equation}
Q({\rm H}^{0}) = \frac{4\pi D^{2}\, F_{\rm obs}\, \alpha_{B}}{h\nu\, \alpha_{\rm eff}(n_{u} \rightarrow n_{l})}.
\end{equation}

Using this formulation, together with the measured line fluxes of all
recombination lines within each of the H\,\textsc{ii} regions and the
recombination coefficients from \citet{storey1987recombination}, we
derived the average ionizing photon production rates for the M33
H\,\textsc{ii} regions in our sample, obtained by averaging the values
inferred from all detected hydrogen recombination lines within each
region. In calculating these values, we explicitly adopt an electron temperature of $T_{e}=10^{4}$~K and an electron density of $n_{e}=100~\mathrm{cm^{-3}}$, consistent with the physical conditions assumed in \citet{maragkoudakis2018pahs}. These parameters also fall comfortably within the validity range of the Case~B calculations of \citet{storey1987recombination} ($10^{3}\,{\rm K} \leq T_{e} \leq 10^{4.5}\,{\rm K}$; $10^{2}\,{\rm cm^{-3}} \leq n_{e} \leq 10^{10}\,{\rm cm^{-3}}$).

Our inferred Lyman-continuum photon production rates given in Table~\ref{table: HII properties}, spanning $Q({\rm H}^{0}) \sim 10^{49}$--$10^{50}\,{\rm s^{-1}}$,
are consistent with expectations for H\,\textsc{ii} regions ionized by early-type O stars
\citep{panagia1973some}, lending confidence to the adopted methodology and the inferred physical conditions of the M33 nebulae. We note, however, that these estimates are derived under the assumption of isotropic emission. If the ionized gas exhibits anisotropic emission,
the inferred $Q({\rm H}^{0})$ values should be regarded as \emph{lower limits} along the line of sight, as directional confinement of the ionizing radiation would enhance the observed recombination-line flux relative to the isotropic case. We further note that the moderate inclination of M33 introduces only projection effects on the observed solid angle and is therefore not expected to produce variations at the order-of-magnitude level in the inferred fluxes.

\subsection{Ionization Index and Ionic Abundance Analysis}
\label{section: Ionization Index and Ionic Abundance Analysis}

\begin{figure*}
\begin{center}
\resizebox{16.0cm}{13.0cm}{\includegraphics{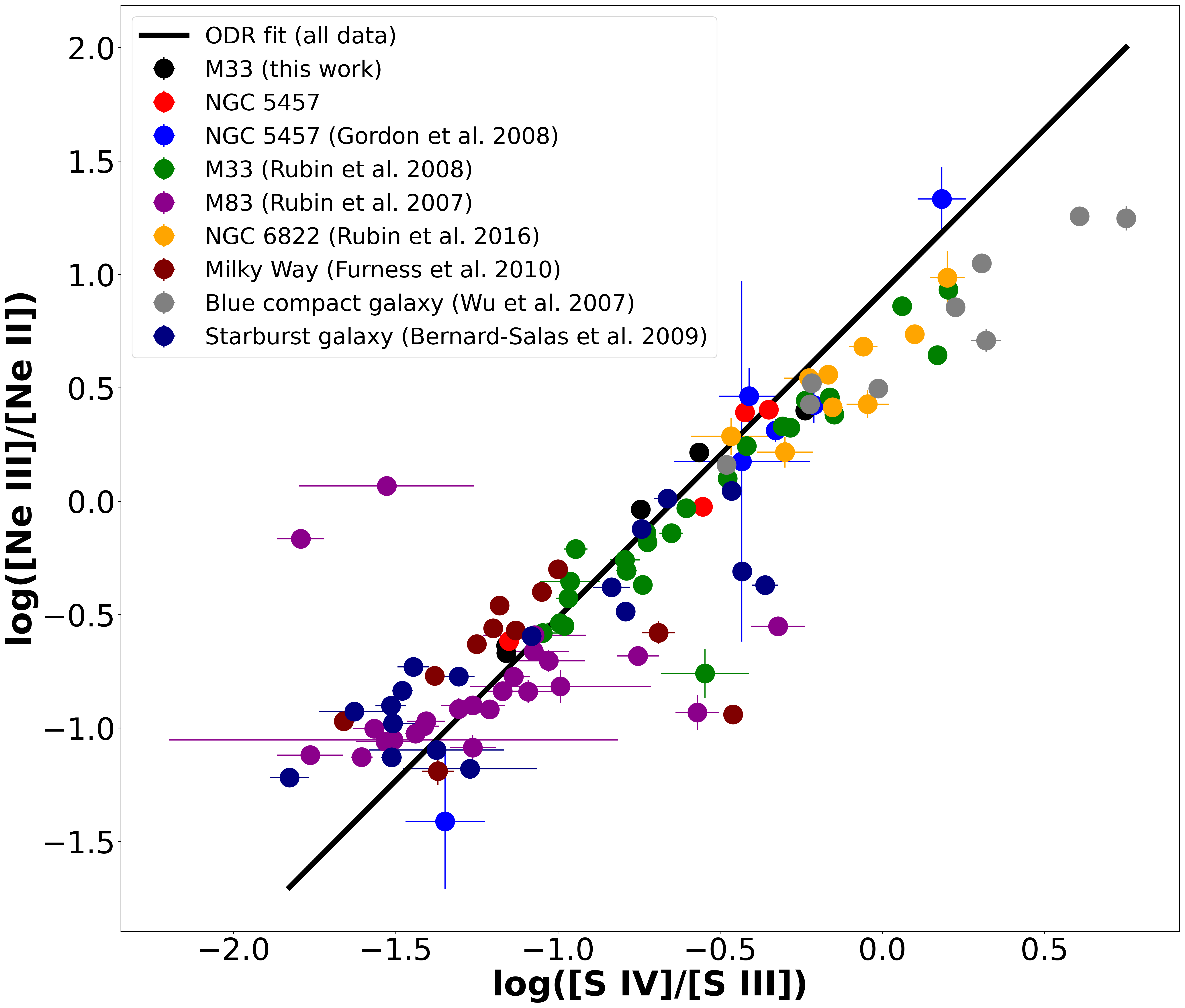}}
\caption{Relative ionic line ratios, $\log([\mathrm{Ne\,\textsc{iii}}]/[\mathrm{Ne\,\textsc{ii}}])$ versus $\log([\mathrm{S\,\textsc{iv}}]/[\mathrm{S\,\textsc{iii}}])$, for the five H\,\textsc{II} regions in M33 analyzed in this work, supplemented by measurements compiled from previous \textit{Spitzer}-based studies of H\,\textsc{ii} regions and other galaxies \citep{gordon2008behavior,rubin2008spitzer,rubin2007spitzer,rubin2016spitzer,furness2010mid,wu2007elemental,bernard2009spitzer}. The sample also includes four H\,\textsc{ii} regions in NGC~5457 observed with JWST/MIRI MRS, for which the integrated ionic line fluxes were measured in this work using the same analysis procedure adopted for the M33 regions. The solid black line represents the best-fitting ODR relation used to calibrate the ionization index.}
\label{Fig: ionization index image}
\end{center}
\end{figure*}

Using the JWST/MIRI observations of the five H\,\textsc{ii} regions in M33, we detect several prominent mid-infrared fine-structure lines from argon, neon and sulfur, including [Ar\,\textsc{ii}], [Ne\,\textsc{ii}], [Ne\,\textsc{iii}], [S\,\textsc{iii}], and [S\,\textsc{iv}]. The ionization potentials associated with the relevant ionic stages for [Ne\,\textsc{ii}], [Ne\,\textsc{iii}], [S\,\textsc{iii}], and [S\,\textsc{iv}]---40.96~eV, 63.45~eV, 34.79~eV, and 47.22~eV, respectively---are taken from the National Institute of Standards and Technology (NIST) Atomic Spectra Database. The fluxes of these lines provide sensitive diagnostics of the ionization state of the nebular gas and allow us to quantify the hardness of the ionizing radiation field. The \emph{ionization index} (II), introduced by \citet{gordon2008behavior}, combines two independent ionic ratios—[Ne\,\textsc{iii}]/[Ne\,\textsc{ii}] and [S\,\textsc{iv}]/[S\,\textsc{iii}]—that respond to different segments of the extreme–UV stellar continuum. Hard UV photons from massive OB stars differentially populate successive ionization stages of neon and sulfur; as a result, these mid–IR fine–structure lines may serve as robust tracers of the spectral energy distribution of the ionizing stellar population.

The ratio [Ne\,\textsc{iii}]/[Ne\,\textsc{ii}] is especially sensitive to photons with energies $\gtrsim 63$ eV, whereas [S\,\textsc{iv}]/[S\,\textsc{iii}] traces comparatively softer photons with ionization potentials above $\sim 47$ eV. Although each ratio alone correlates with the hardness of the radiation field, both can be affected by local ionization structure, metallicity gradients, and geometric projection effects. Empirical studies of H\,\textsc{ii} regions in nearby galaxies, particularly M101, reveal a remarkably tight relationship between $\log([\mathrm{Ne\,III}]/[\mathrm{Ne\,II}])$ and $\log([\mathrm{S\,IV}]/[\mathrm{S\,III}])$ \citep{gordon2008behavior}. This correlation arises because both ionic pairs are regulated by similar underlying ionizing continua and therefore respond coherently to changes in the effective temperature of the dominant ionizing sources. \citet{engelbracht2008metallicity} demonstrated that this relationship can be linearized, enabling the sulfur ratio to be translated onto a “neon-equivalent” scale and yielding a combined, more stable tracer of the nebular ionization state.

Following the methodology established in previous studies \citep{gordon2008behavior, maragkoudakis2018pahs}, we derived an empirical calibration between $\log([\mathrm{Ne\,\textsc{iii}}]/[\mathrm{Ne\,\textsc{ii}}])$ and $\log([\mathrm{S\,\textsc{iv}}]/[\mathrm{S\,\textsc{iii}}])$ to estimate the ionization index for the five H\,\textsc{ii} regions in M33 observed with JWST/MIRI MRS. To construct a statistically robust calibration spanning a wide range of ionization conditions and galactic environments, we complemented our measurements with published mid-infrared ionic line fluxes from \textit{Spitzer} observations of H\,\textsc{ii} regions in M33 \citep{rubin2008spitzer}, M83 \citep{rubin2007spitzer}, M101 (NGC 5457) \citep{gordon2008behavior}, and NGC~6822 \citep{rubin2016spitzer}, as well as the Galactic giant star-forming region W31 \citep{furness2010mid}, blue compact dwarf galaxies \citep{wu2007elemental}, and starburst galaxies \citep{bernard2009spitzer}. In addition, we retrieved JWST/MIRI MRS observations of four H\,\textsc{ii} regions in NGC~5457 from the MAST archive. Although these regions were previously investigated by \citet{singh2025jwst} primarily for their PAH characteristics, we independently measured the integrated fluxes of the relevant ionic emission lines using the same analysis procedure adopted for our M33 sample, thereby ensuring a homogeneous treatment of the entire dataset.

The resulting compilation was used to examine the relationship between $\log([\mathrm{Ne\,\textsc{iii}}]/[\mathrm{Ne\,\textsc{ii}}])$ and $\log([\mathrm{S\,\textsc{iv}}]/[\mathrm{S\,\textsc{iii}}])$. As illustrated in Figure~\ref{Fig: ionization index image}, the diagnostic exhibits a strong and well-defined linear correlation, with a Pearson correlation coefficient of $\sim0.90$ and an RMS scatter of $\sigma_{Y/X}=0.38$. The tightness of this relation demonstrates that both ionic ratios respond in a similar manner to the hardness of the ionizing radiation field across a diverse sample of H\,\textsc{ii} regions and star-forming galaxies.

To account for measurement uncertainties in both variables, we performed an orthogonal distance regression (ODR) analysis and obtained the following best-fitting relation:
\begin{equation} \begin{split} \log\!\left(\frac{\mathrm{[Ne\,III]}}{\mathrm{[Ne\,II]}}\right) &= (0.922 \pm 0.025) \\ &\quad + (1.436 \pm 0.055) \log\!\left(\frac{\mathrm{[S\,IV]}}{\mathrm{[S\,III]}}\right), 
\end{split}
\label{equation:neon_to_sulfur_flux}
\end{equation}
Using this calibration, the ionization index is defined as the mean of the observed neon ratio and its sulfur-based proxy:
\begin{equation}
    \mathrm{II} = \frac{1}{2}\left[
    \log\!\left(\frac{\mathrm{[Ne\,III]}}{\mathrm{[Ne\,II]}}\right)
    + 0.922 + 1.436
      \log\!\left(\frac{\mathrm{[S\,IV]}}{\mathrm{[S\,III]}}\right)
    \right].
    \label{equation: ionization index equation}
\end{equation}

Using Equation~\ref{equation: ionization index equation}, we compute the ionization index for each of the five H\,\textsc{ii} regions, as reported in Table~\ref{table: HII properties}. Among the sample, M33–438+800 exhibits the highest II value, indicating a particularly hard radiation field consistent with the presence of hotter or more massive ionizing stars. Two regions, M33+553+448 and M33–72–1072, show modestly positive II values, whereas the remaining two regions have negative II values, implying significantly softer radiation fields and possibly more evolved or lower–mass ionizing stellar populations. The implications of these ionization conditions for PAH processing is examined in detail in Section~\ref{section: PAH Ionization Diagnostics: Trends and Interpretation}.

In addition to constraining the ionization state, the same metallic fine–structure lines allow us to derive the ionic abundances of neon and sulfur. For two ionic species $X^{+i}$ and $X^{+j}$, their abundance ratio is related to the observed flux ratio by
\begin{equation}
    \frac{X^{+i}}{X^{+j}}
    = \frac{F(X^{+i})}{F(X^{+j})}
      \frac{\epsilon_{X^{+j}}}{\epsilon_{X^{+i}}},
    \label{equation: ionic abundance}
\end{equation}
where $\epsilon$ is the emissivity coefficient of the transition. Emissivity values are computed using the \textsc{PyNeb} package \citep{luridiana2015pyneb}, a widely used Python tool for modeling ionized nebulae that incorporates the most up–to–date atomic data for collisional excitation, effective recombination coefficients, and radiative decay rates. We adopt $T_{e}=8000$~K and $n_{e}=100~\mathrm{cm^{-3}}$, following the physical conditions reported by \citet{maragkoudakis2018pahs} for similar M33 H\,\textsc{ii} regions. Columns 7-12 of Table~\ref{table: HII properties} present the derived ionic abundance ratios—Ne$^{+2}$/Ne$^{+}$, S$^{+3}$/S$^{+2}$, Ne$^{+2}$/S$^{+2}$, Ne$^{+2}$/S$^{+3}$, Ne$^{+}$/S$^{+2}$, and Ne$^{+}$/S$^{+3}$—for all five regions. The dependence of these abundance ratios on the ionization index is discussed in Section~\ref{section: ionic_abundance_variation}.

Following the methodology adopted by \cite{maragkoudakis2018pahs}, we estimated the sulfur-based metallicity of the H\,\textsc{ii} regions using the mid-infrared fine-structure emission lines of [S\,\textsc{iii}] and [S\,\textsc{iv}] together with the H\,I (7--6) recombination line (Hu\,$\alpha$). Assuming that S$^{2+}$ and S$^{3+}$ are the dominant ionization stages of sulfur in the ionized gas, we first derived the ionic abundances S$^{2+}$/H$^{+}$ and S$^{3+}$/H$^{+}$ from the observed line-to-recombination-line flux ratios using the corresponding emissivities at the adopted electron temperature, T$_{e}$ = 8000 K and density, n$_{e}$ = 100 cm$^{-2}$. The total sulfur abundance was then obtained by summing these two ionic components and expressed in the conventional logarithmic form as $12+\log(\mathrm{S}/\mathrm{H})$.

To facilitate comparison with the solar composition, the sulfur abundance was converted into a metallicity relative to the solar value using the relation
$Z/Z_{\odot} = 10^{\left[(12+\log(\mathrm{S}/\mathrm{H})) - (12+\log(\mathrm{S}/\mathrm{H})_{\odot})\right]}$,
where we adopted the solar sulfur abundance
$12+\log(\mathrm{S}/\mathrm{H})_{\odot}=7.12$
from \cite{asplund2009chemical}. The resulting sulfur-based metallicities, expressed in units of $Z_{\odot}$, are listed in Table~\ref{table: HII properties} for all five H\,\textsc{ii} regions in M33. We found that all five H\,\textsc{ii} regions exhibit subsolar metallicities, consistent with the well-established picture of M33 as a relatively low-metallicity spiral galaxy and in agreement with previous studies \citep{maragkoudakis2018pahs,rosolowsky2008m33}.

We adopt different electron temperatures for different diagnostics, following common practice and previous studies of M33 H\,\textsc{ii} regions. The ionizing photon rate $Q(\mathrm{H}^0)$ was computed assuming $T_e = 10^4$~K under Case~B recombination, consistent with standard recombination coefficients and with the approach adopted by \citet{maragkoudakis2018pahs}. In contrast, ionic abundances derived using \texttt{PyNeb} were computed assuming $T_e = 8000$~K, which is more representative of the characteristic nebular temperatures measured in M33 H\,\textsc{ii} regions \citep{maragkoudakis2018pahs}. An electron density of $n_e = 100~\mathrm{cm^{-3}}$ was adopted throughout, typical of classical H\,\textsc{ii} regions and within the low-density regime where the relevant mid-infrared line emissivities are only weakly dependent on density. We verified that varying $T_e$ within the range $7000$--$12{,}000$~K does not affect our main conclusions.

\section{Discussion} 
\label{section: discussion}

\subsection{PAH Ionization Diagnostics: Trends and Interpretation}
\label{section: PAH Ionization Diagnostics: Trends and Interpretation}

\begin{figure*}
\begin{center}
\resizebox{12.0cm}{10.0cm}{\includegraphics{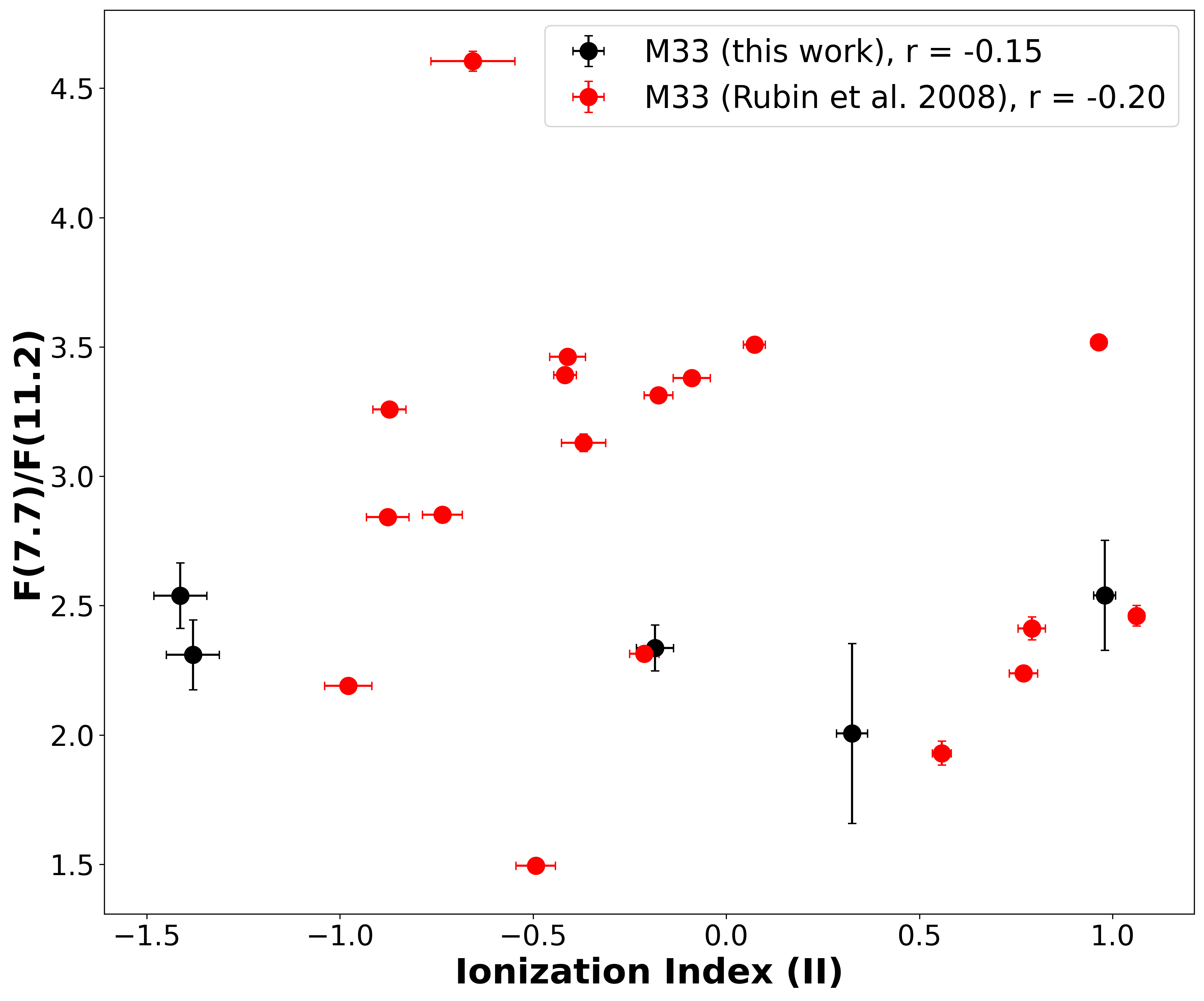}}
\caption{Dependence of the PAH flux ratios F(7.7)/F(11.2) on the ionization index.}\label{Fig: PAH_ionization image}
\end{center}
\end{figure*}

The JWST/MIRI observations reveal three prominent major PAH emission bands at 7.7, 8.6, and 11.2~$\mu$m, together with a relatively weaker 11.0~$\mu$m feature, in all five H\,\textsc{ii} regions in M33. In addition, a weak 16.4~$\mu$m PAH feature is detected in all regions except M33-72-1072. The 7.7~$\mu$m complex exhibits a Class A profile, with its peak centred near 7.6~$\mu$m \citep{tielens2005physics}. Such a spectral profile is characteristic of PAH emission in H\,\textsc{ii} regions and PDRs \citep{tielens2005physics}. Similar Class A profiles have been reported in numerous Galactic and extragalactic star-forming environments, including 30 Doradus \citep{zhang2025jwst}, the Orion Bar \citep{chown2024pdrs4all}, H\,\textsc{ii} regions in NGC~5457 \citep{singh2025jwst}, and H\,\textsc{ii} regions in M33 and M83 observed with \textit{Spitzer} \citep{maragkoudakis2018pahs}. These similarities indicate that the PAH spectral characteristics of our sample are consistent with those commonly observed in massive star-forming environments.

To place our results into the broader context of PAH studies in M33, we compare the derived PAH properties of our five JWST-observed H\,\textsc{ii} regions with the sample of 18 H\,\textsc{ii} regions analysed using \textit{Spitzer} spectroscopy by \citet{maragkoudakis2018pahs}. Although one of our targets, M33+62+354, appears to spatially coincide with Region~301 from their catalogue (Figure~\ref{Fig: UV image of M33 galaxy}), this apparent correspondence primarily results from the differing spatial resolutions of the observations. The GALEX ultraviolet image used for source visualization has an angular resolution of approximately 5.3 arcsec at 227~nm, whereas the projected separation between the centres of M33+62+354 and Region~301 is only 4.42 arcsec (corresponding to $\sim$18~pc at a distance of 840~kpc), which is smaller than the GALEX resolution. Furthermore, \citet{maragkoudakis2018pahs} employed an aperture of approximately 104.04~arcsec for the 5.1--39.9~$\mu$m wavelength range, encompassing a substantially larger area than the JWST observations. Consequently, the JWST source M33+62+354 represents only a small substructure within the much larger \textit{Spitzer} Region~301. Owing to these significant differences in spatial resolution and extraction aperture, a direct one-to-one comparison of the derived physical properties is not appropriate. Instead, we compare our PAH properties statistically with the full sample of 18 H\,\textsc{ii} regions presented by \citet{maragkoudakis2018pahs}.

The mean 7.7/11.2 PAH intensity ratio derived for our sample is 2.35, which is lower than the average value of 3.09 reported by \citet{maragkoudakis2018pahs}. This result further supports the trend that the 7.7/11.2 PAH ratio decreases with decreasing physical scale. \citet{maragkoudakis2018pahs} demonstrated that the average 7.7/11.2 PAH ratio measured in individual H\,\textsc{ii} regions of M33 is systematically lower than that observed in integrated Spitzer
Infrared Nearby Galaxies Survey (SINGS) H\,\textsc{ii}-type galaxies, attributing this difference to the transition from galaxy-scale measurements to individual giant star-forming regions. Our results extend this trend to even smaller spatial scales. Taking advantage of the superior spatial resolution of \textit{JWST}, we derive systematically lower average 7.7/11.2 PAH ratios than those obtained from the lower-resolution \textit{Spitzer} observations of the same galaxy. This decrease is most likely driven by the smaller extraction apertures enabled by \textit{JWST}, which isolate localized physical environments that are spatially diluted within the larger \textit{Spitzer} apertures. Although the PAH fluxes in our analysis are integrated over a broader wavelength range than those adopted by \citet{maragkoudakis2018pahs}, the observed reduction in the mean 7.7/11.2 PAH ratio remains consistent with the proposed scale dependence. These findings therefore reinforce the conclusion that the 7.7/11.2 PAH intensity ratio systematically declines as observations probe progressively smaller physical scales, reflecting variations in the local physical conditions that regulate PAH emission.

In the study of \cite{maragkoudakis2018pahs}, the 7.7/11.2 and 8.6/11.2 PAH intensity ratios exhibit a strong positive correlation, where the 7.7 and 11.2~$\mu$m bands were defined over wavelength intervals of 7.3--7.9~$\mu$m and 11.2--11.4~$\mu$m, respectively, using \textsc{PAHFIT}. In contrast, we do not find statistically significant evidence for a correlation between these two ratios within our sample, despite both being widely used as tracers of the PAH ionization fraction, since the 7.7 and 8.6~$\mu$m bands predominantly arise from cationic PAHs, whereas the 11.2~$\mu$m feature is primarily associated with neutral PAHs. This discrepancy is likely attributable to two factors. First, the complete 8.6~$\mu$m feature is not covered by the JWST/MIRI observations used in this work owing to the instrumental wavelength limitation discussed in Section~\ref{section: Flux Measurement of PAH Bands and Narrow Emission Lines}. Second, our sample consists of only five H\,\textsc{ii} regions, limiting the statistical significance of any underlying correlation.

\citet{maragkoudakis2018pahs} further showed that most H\,\textsc{ii} regions in M33 occupy the regime with [Ne\,\textsc{iii}]/[Ne\,\textsc{ii}]~$\lesssim$~1, corresponding to relatively soft radiation fields. All five H\,\textsc{ii} regions in our sample likewise exhibit [Ne\,\textsc{iii}]/[Ne\,\textsc{ii}]~$<1$, suggesting that they probe a similar radiation-field regime and do not sample the harder ionizing environments where substantial PAH processing is expected. Consistent with the findings of \citet{maragkoudakis2018pahs}, we do not find statistically significant evidence for a correlation between the 7.7/11.2 ratio and either the ionization index (Figure~\ref{Fig: PAH_ionization image}) or the [Ne\,\textsc{iii}]/[Ne\,\textsc{ii}] ratio.

We observe an apparent decrease in the 7.7/11.2 PAH intensity ratio with increasing sulphur metallicity, with a Pearson correlation coefficient of $r=-0.61$ ($p=0.30$, $N=5$). Although this suggests a possible anti-correlation, the relationship is not statistically significant owing to the limited sample size. This behavior contrasts with the results of \citet{maragkoudakis2018pahs}, who reported no clear dependence of the 7.7/11.2 PAH ratio on metallicity. The discrepancy may arise from several factors, including the substantially different spatial scales probed by \textit{JWST} and \textit{Spitzer}, the smaller extraction apertures afforded by the superior spatial resolution of \textit{JWST}, and differences in the adopted wavelength integration ranges used to measure the PAH band fluxes. Consequently, it remains uncertain whether the observed trend reflects a genuine metallicity dependence or is primarily driven by observational and methodological differences. 

Very small grains (VSGs; \citealt{desert1990interstellar}) are generally believed to consist of carbonaceous particles only a few nanometres in size. These grains are stochastically heated by ultraviolet photons within H\,\textsc{ii} regions and emit predominantly at wavelengths longer than $\sim10~\mu$m. Following \citet{madden2006ism} and \citet{maragkoudakis2018pahs}, we define the VSG emission as the integrated dust continuum between 10 and 16~$\mu$m. However, unlike \citet{maragkoudakis2018pahs}, who defined the total PAH emission as the sum of the 6.2, 7.7, 8.6, 11.3, and 12.6~$\mu$m bands, our definition is necessarily modified because the 6.2~$\mu$m feature lies outside the observed wavelength coverage, the 16.4~$\mu$m feature is not detected in all regions, and the common PAH bands measured for every source are the 7.7, 8.6, 11.0, and 11.2~$\mu$m features. Using this definition, we find that the total PAH-to-VSG emission ratio decreases with increasing radiation hardness, as traced by [Ne\,\textsc{iii}]/[Ne\,\textsc{ii}], consistent with the trends reported by \citet{madden2006ism} and subsequently confirmed for the H\,\textsc{ii} regions in M33 by \citet{maragkoudakis2018pahs}. This behaviour supports the interpretation that progressively harder radiation fields preferentially destroy PAH molecules relative to the more resilient VSG population.

Although the superior spatial resolution and sensitivity of JWST provide an unprecedented view of the PAH emission within individual H\,\textsc{ii} regions, the conclusions drawn here remain limited by the small sample size. Consequently, the relationships involving the 7.7/11.2 ratio with the 8.6/11.2 ratio, ionization index, [Ne\,\textsc{iii}]/[Ne\,\textsc{ii}], and metallicity dependence should be regarded as preliminary. Future JWST observations targeting the 18 H\,\textsc{ii} regions studied by \citet{maragkoudakis2018pahs}, together with additional H\,\textsc{ii} regions throughout M33, will enable a much more robust statistical investigation of the dependence of PAH properties on the local physical conditions within the galaxy by minimizing the aperture dilution effects inherent in lower-resolution \textit{Spitzer} observations.

Following the methodology of \citet{zhang2025jwst}, we derived several physical properties of the PAHs in our sample of five H\,\textsc{ii} regions, including the PAH absorption energy, charge state, and PAH ionization parameter. To estimate the PAH absorption energy, we adopted circumcoronene (C$_{54}$H$_{18}$) as the representative PAH molecule and employed the absorption cross section of neutral circumcoronene from \citet{malloci2007line}, available through the INAF (National Institute for Astrophysics) Astrochemistry Database (http://astrochemistry.ca.astro.it/database/). The absorption energy was calculated for far-ultraviolet photons following Equation~(1) of \citet{croiset2016mapping}, assuming an extinction curve corresponding to $R_{V}=5.5$. For the FUV extinction, we adopted the extinction law of \citet{valencic2004ultraviolet}. The total hydrogen column density was estimated as $N(\mathrm{H})=2\times N(\mathrm{H}_{2})$, and the extinction correction was computed using $\tau=A_{\lambda}/1.086$ and $e^{-\tau}$.

Among the five H\,\textsc{ii} regions, probable O-type ionizing stars have been identified for four regions (excluding M33-72-1072; \citealt{massey1996uv, massey2006survey}). However, the spectral types of these ionizing sources are not available in the literature. Therefore, to estimate the PAH absorption energy, we considered a representative stellar effective temperature range of $30\,000$--$50\,000$~K. We find that the derived PAH absorption energies vary only weakly across this temperature range, yielding a mean value of approximately 6.17~eV with a fractional variation of only 1.21\%. This weak dependence indicates that the absorption energy is relatively insensitive to the assumed stellar temperature within the expected range for O-type stars, thereby supporting the robustness of the derived values.

The PAH charge state was subsequently estimated following the diagnostic method of \citet{zhang2025jwst}, based on the charge--size grids presented by \citet{maragkoudakis2020probing}. For the four H\,\textsc{II} regions with estimated absorption energies, we first computed the intensity ratio $(F_{11.0}+F_{11.2})/F_{7.7}$, obtaining values of 0.43, 0.47, 0.46, and 0.42 for M33-438+800, M33+553+448, M33+46-380, and M33+62+354, respectively. Using the charge--size grid corresponding to an absorption energy of $\sim$6~eV, these ratios indicate that approximately 75\% of the PAH population in all four H\,\textsc{ii} regions is in the cationic state.

It should be noted, however, that the determination of both PAH charge state and size ideally requires the inclusion of the 3.3~$\mu$m PAH feature. As the JWST/MIRI observations analysed in this work do not include NIRSpec IFU spectroscopy, the 3.3~$\mu$m band is unavailable, preventing a direct constraint on the PAH size distribution. Future JWST/NIRSpec IFU observations of these regions will therefore be crucial for simultaneously constraining the PAH size distribution and charge state, while also providing valuable insights into the formation and destruction mechanisms of PAH molecules within the M33 galaxy.

Interestingly, despite the relatively soft radiation fields inferred from the observed [Ne\,\textsc{iii}/[Ne\,\textsc{ii}] ratios ($<1$; \citealt{maragkoudakis2018pahs}), the diagnostic analysis indicates that the PAH population in all four H\,\textsc{ii} regions is nevertheless predominantly cationic. This result suggests that even moderate ultraviolet radiation fields in these star-forming environments are sufficient to maintain a largely ionized PAH population, consistent with the strong 7.7~$\mu$m emission characteristic of cationic PAHs.

The probable ionizing stars associated with four of our H\,\textsc{ii} regions (M33-438+800, M33+553+448, M33+46-380, and M33+62+354) were identified by \citet{massey1996uv} as UIT092 (RA = 01:33:16.55, Dec = +30:52:49.6), UIT368 (RA = 01:34:33.44, Dec = +30:46:57.1), UIT240 (RA = 01:33:54.09, Dec = +30:33:09.9), and UIT244 (RA = 01:33:55.47, Dec = +30:45:24.0), respectively. Among these sources, only UIT240 has a spectroscopic classification (O6--8If), whereas the remaining three objects are simply classified as H\,\textsc{ii} sources by \citet{massey1996uv}. Consequently, the spectral types of the ionizing stars remain uncertain for three of the four regions.

To place approximate constraints on the nature of the ionizing sources, we considered the measured Lyman continuum photon production rates, $Q(\mathrm{H}^{0})$, listed in Table~\ref{table: HII properties}. The four regions have $\log Q(\mathrm{H}^{0}) \approx 50$, 51, 50, and 50, respectively. Stellar atmosphere models indicate that such ionizing photon fluxes are broadly consistent with massive early O-type stars on or close to the zero-age main sequence (ZAMS) or main-sequence phase, whereas B-type stars typically produce significantly lower ionizing photon rates ($\log Q(\mathrm{H}^{0}) \sim 44$--47; \citealt{panagia1973some,stahler2008formation}). Since no additional spectral classifications are available from the \textit{Gaia} mission \citep{vallenari2023gaia}, we adopt early O-type stars as a reasonable first-order approximation for estimating the incident radiation field. We emphasize, however, that this interpretation is not unique, as the observed ionizing photon flux could equally arise from multiple late O-type stars or a compact cluster containing several B-type stars.

The projected separations between the identified ionizing stars and the locations of maximum PAH emission are 1.58, 1.66, 5.56, and 5.69~pc for M33-438+800, M33+553+448, M33+46-380, and M33+62+354, respectively. Assuming isotropic radiation from the ionizing sources and adopting the approximation that, for early O-type stars, the far-ultraviolet ($6 < h\nu < 13.6$~eV) photon flux is comparable to the Lyman continuum photon flux \citep{stahler2008formation}, we estimated the incident FUV radiation field in units of the Habing field ($G_{0}$; $1.6 \times 10^{-3}$ erg cm$^{-2}$ s$^{-1}$; \citealt{habing1968interstellar}). 


Using the $G_{0}$ values, we calculated the PAH ionization parameter \citep{berne2022contribution, boersma2016charge, zhang2025jwst},
\begin{equation}
\gamma=\frac{G_{0}\sqrt{T_{\rm gas}}}{n_{e}},
\end{equation}
where $T_{\rm gas}$ is the gas temperature and $n_{e}$ is the electron density in the photodissociation region. The rotational temperature derived from the warm H$_2$ emission was adopted as a proxy for the PDR gas temperature, while the PDR electron density was assumed to lie within the range (2.5 $\times 10^{-3}$ - 75 cm$^{-3}$) proposed by \citet{bakes1994photoelectric}. Since the electron density in PDR is not directly constrained by the JWST/MIRI observations used in our work, we evaluated $\gamma$ over the full range of plausible $n_{e}$ values.


Table~\ref{table:G0 gamma} summarizes the derived range of the incident far-ultraviolet radiation field strength, $G_0$, and the corresponding values of the ionization parameter, $\gamma$. The $\gamma$ values are evaluated separately by considering the lower and upper limits of the electron density, $n_{\rm e}$, within the PDR, and are denoted as $\gamma^{\rm upper}$ and $\gamma^{\rm lower}$, respectively.

Our PAH spectral analysis indicates that nearly 75\% of the PAH population in these H\,\textsc{ii} regions is in the cationic state. Theoretical studies have shown that environments with predominantly ionized PAHs are generally associated with relatively large values of the PAH ionization parameter \citep{boersma2016charge, berne2022contribution}. Although the PAH charge state is governed not only by $\gamma$ but also by the PAH size distribution, molecular structure, absorbed photon energy, and other local physical conditions, the predominance of cationic PAHs inferred from the JWST/MIRI observations analysed in our work is qualitatively more consistent with the higher-$\gamma$ regime than with the lowest values obtained by adopting the maximum PDR electron density. This comparison therefore suggests that the actual electron densities within the unresolved PDRs are likely to be lower than the upper limit adopted from \citet{bakes1994photoelectric}. However, this interpretation should be regarded as indicative rather than conclusive, since neither the PDR electron density nor the FUV radiation field within PDR regions is directly measured in the present work. Future spatially resolved observations capable of independently constraining the physical conditions within the PDRs will be essential for deriving more accurate values of the PAH ionization parameter.

We further compared the derived values of $\gamma$ with the observed 7.7/11.2 PAH intensity ratios to examine whether the PAH ionization parameter directly traces the PAH charge state. No clear one-to-one correspondence is found between these two quantities. This is not unexpected because, although both diagnostics are related to the PAH ionization state, the 7.7/11.2 ratio is also affected by several additional factors, including the PAH size distribution, molecular structure, absorbed photon energy, and local physical conditions. Consequently, the absence of a tight correlation does not imply that $\gamma$ is unrelated to the PAH charge state; rather, it reflects the intrinsically multi-parameter nature of PAH emission together with the relatively limited dynamic range and small sample size of the five H\,\textsc{ii} regions investigated in this work.

Finally, we emphasize that the derived values of both $G_{0}$ and $\gamma$ should be regarded as first-order estimates rather than precise measurements. Their uncertainties arise from several simplifying assumptions, including the unknown spectral types of most ionizing stars, the approximation that the FUV photon flux is comparable to the ionizing photon flux, the adoption of LTE rotational temperatures as proxies for the PDR gas temperature, and the use of an approximate PDR electron density following \citet{bakes1994photoelectric} instead of a direct observational determination. Furthermore, the PDRs surrounding these H\,\textsc{ii} regions are unresolved in the present observations, preventing a self-consistent determination of their physical conditions. Nevertheless, the estimated values of $G_{0}$ and $\gamma$ provide a physically motivated first-order characterization of the radiation field and PAH ionization conditions in these H\,\textsc{ii} regions and offer a useful framework for interpreting the observed PAH emission.

\begin{table}[ht]
\centering
\caption{Derived values of $G_0$ and $\gamma$ for the five M33 H\,\textsc{ii} regions.}
\label{table:G0 gamma}

\begin{tabular}{ccccccc}
\hline
Index & \multicolumn{2}{c}{$G_0 \times 10^{3}$} & \multicolumn{2}{c}{$\gamma^{upper} \times 10^{5}$} & \multicolumn{2}{c}{$\gamma^{lower} \times 10^{3}$} \\
\cline{2-3} \cline{4-5} \cline{6-7}
& Min & Max & Min & Max & Min & Max \\
\hline
1 & 6.92 & 15.69 & 492 & 1116 & 1.64 & 3.72\\
2 & 10.77 & 24.42 & 838 & 1900 & 2.80 & 6.34\\
3 & 0.72 & 1.64 & 58 & 132 & 0.19 & 0.44\\
4 & 0.33 & 0.74 & 25 & 58 & 0.09 & 0.19\\

\hline
\end{tabular}

\end{table}

\subsection{Variation of Ionic Abundances with the Ionization Index}
\label{section: ionic_abundance_variation}

\begin{figure*}
\begin{center}
\resizebox{18.0cm}{11.5cm}{\includegraphics{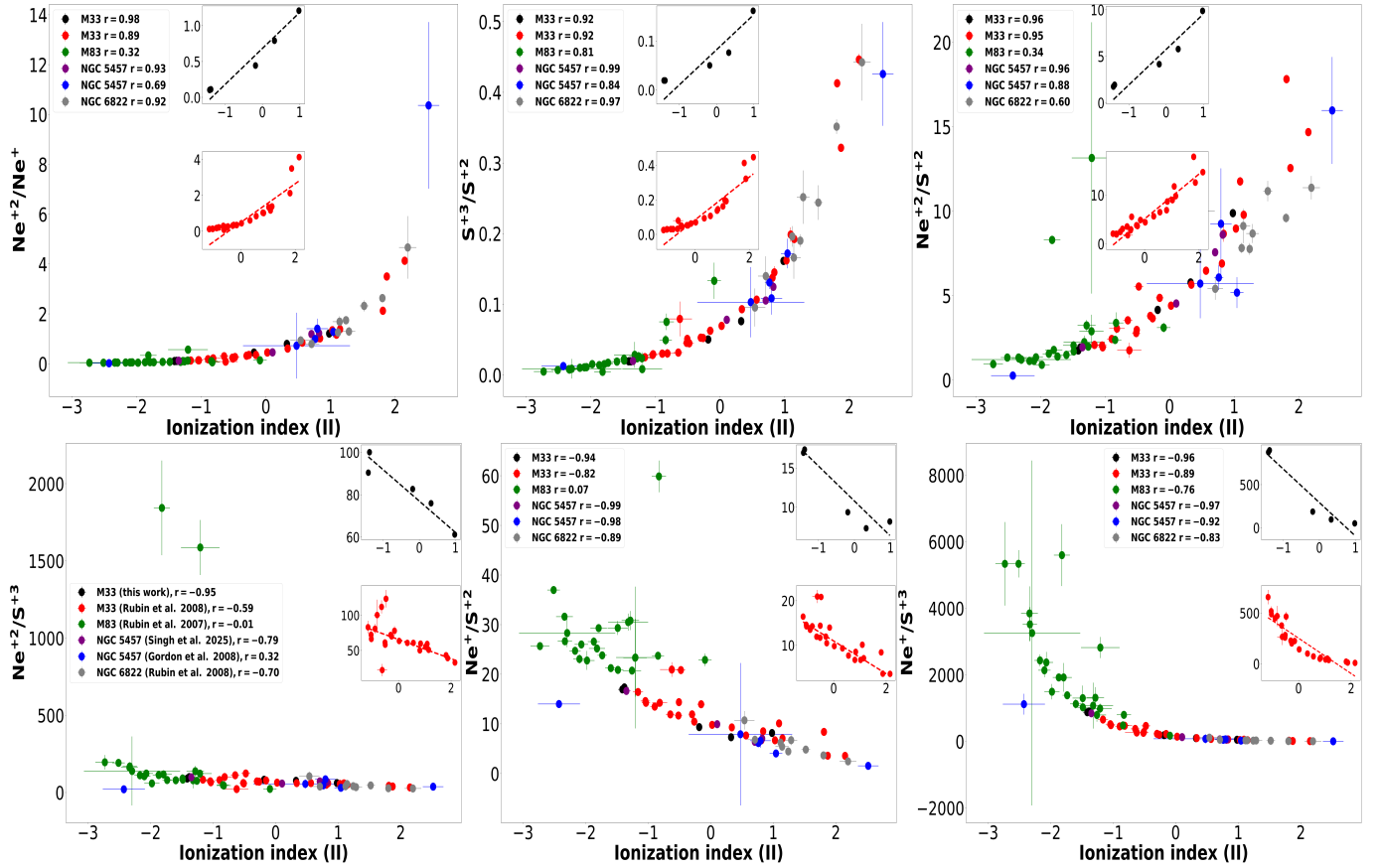}}
\caption{Variation of the ionic abundance ratios as a function of the ionization index (II). The upper inset shows the H,\textsc{ii} regions in M33 analyzed in this work using archival \textit{JWST}/MIRI observations, while the lower inset presents the corresponding M33 H,\textsc{ii} regions studied by \citet{rubin2008spitzer} using \textit{Spitzer} observations. The black and red dashed lines in the upper and lower insets, respectively, represent the best-fitting linear relations to the corresponding data sets, illustrating the observed dependence of the ionic abundance ratios on the ionization state of the ionized gas. The black, red, green, purple, blue, and gray symbols represent the measurements for M33 from this work, M33 from \citet{rubin2008spitzer}, M83 from \citet{rubin2007spitzer}, NGC~5457 from \citet{singh2025jwst}, NGC~5457 from \citet{gordon2008behavior}, and NGC~6822 from \citet{rubin2016spitzer}, respectively. For NGC~5457  studied in \cite{singh2025jwst}, the ionic abundance ratios and II values are derived from archival \textit{JWST}/MIRI data. The strength of the linear correlation for each galaxy sample is quantified by the Pearson correlation coefficient, $r$, with the corresponding values indicated in each panel.
}\label{Fig: IB vs II relation}
\end{center}
\end{figure*}

To investigate how the hardness of the radiation field influences the ionic structure of H\,\textsc{ii} regions, we examined the dependence of several ionic abundance ratios on the ionization index (II). Figure~\ref{Fig: IB vs II relation} compares our JWST measurements for the five H\,\textsc{ii} regions in M33 with previous observations of H\,\textsc{ii} regions in M33 and M83 from \citet{maragkoudakis2018pahs}, H\,\textsc{ii} regions in NGC~5457 from \citet{gordon2008behavior}, and the low-metallicity H\,\textsc{ii} regions in NGC~6822 from \citet{rubin2016spitzer}. We also include the ionic abundances derived for the JWST H\,\textsc{ii} regions in NGC~5457 presented by \citet{singh2025jwst}, although that work primarily focused on PAH properties. Since NGC~5457 exhibits a pronounced radial metallicity gradient, its H\,\textsc{ii} regions span a broad metallicity range, providing an excellent comparison sample. For clarity, the upper inset of Figure~\ref{Fig: IB vs II relation} presents the relations for our five JWST H\,\textsc{ii} regions, while the lower inset shows the corresponding measurements for the M33 H\,\textsc{ii} regions observed with \textit{Spitzer} by \citet{maragkoudakis2018pahs}.

The ionic ratios Ne$^{+2}$/Ne$^{+}$, S$^{+3}$/S$^{+2}$, and Ne$^{+2}$/S$^{+2}$ generally exhibit moderate to strong positive correlations with II for the H\,\textsc{ii} regions in M33, NGC~5457, and NGC~6822. In contrast, the M83 sample shows only weak correlations for Ne$^{+2}$/Ne$^{+}$ and Ne$^{+2}$/S$^{+2}$, although the S$^{+3}$/S$^{+2}$ ratio still displays a strong positive dependence on II. The positive trends might be physically expected because increasing II corresponds to a harder ionizing radiation field capable of producing a larger fraction of high-ionization species. As the number of photons with energies exceeding the ionization potentials of Ne$^{+}$ (40.96~eV) and S$^{+2}$ (34.83~eV) increases, the ionic balance shifts toward Ne$^{+2}$ and S$^{+3}$. Consequently, both Ne$^{+2}$/Ne$^{+}$ and S$^{+3}$/S$^{+2}$ increase with radiation hardness. Likewise, the Ne$^{+2}$/S$^{+2}$ ratio rises because Ne$^{+2}$ becomes progressively more abundant while S$^{+2}$ is simultaneously depleted through further ionization to S$^{+3}$.

Conversely, the ionic ratios Ne$^{+2}$/S$^{+3}$, Ne$^{+}$/S$^{+2}$, and Ne$^{+}$/S$^{+3}$ predominantly show moderate to strong negative correlations with II in the M33, JWST NGC~5457, and NGC~6822 samples. The M83 H\,\textsc{ii} regions exhibit little or no correlation for Ne$^{+2}$/S$^{+3}$ and Ne$^{+}$/S$^{+2}$, while Ne$^{+}$/S$^{+3}$ retains a strong negative trend. The H\,\textsc{ii} regions studied by \citet{gordon2008behavior} in NGC~5457 display a weak positive correlation for Ne$^{+2}$/S$^{+3}$ but strong negative correlations for Ne$^{+}$/S$^{+2}$ and Ne$^{+}$/S$^{+3}$. The anti-correlations might arise from the progressive redistribution of ions toward higher ionization stages in harder radiation fields. As II increases, Ne$^{+}$ is efficiently converted into Ne$^{+2}$, while S$^{+2}$ is successively ionized to S$^{+3}$ and higher ionization stages. Ratios containing lower-ionization species in the numerator and higher-ionization species in the denominator therefore decrease systematically with increasing radiation hardness.

An important outcome of this comparison is the remarkable consistency between the JWST/MIRI  observations used in our work and the earlier \textit{Spitzer} measurements of H\,\textsc{ii} regions within M33. Both datasets exhibit strong positive correlations between II and the ionic ratios Ne$^{+2}$/Ne$^{+}$, S$^{+3}$/S$^{+2}$, and Ne$^{+2}$/S$^{+2}$, together with strong negative correlations for Ne$^{+2}$/S$^{+3}$, Ne$^{+}$/S$^{+2}$, and Ne$^{+}$/S$^{+3}$. The agreement between two independent datasets obtained with different instruments and spatial resolutions suggests that these relationships primarily reflect the underlying ionization physics of M33 H\,\textsc{ii} regions rather than observational biases.

Nevertheless, the comparison among galaxies demonstrates that these correlations are not strictly universal. Although the low-metallicity H\,\textsc{ii} regions in NGC~6822 broadly follow the same behaviour observed in M33, the M83 sample exhibits noticeably weaker or absent correlations for several ionic ratios, and some of the H\,\textsc{ii} regions in NGC~5457 display different trends. Since NGC~5457 spans a wide metallicity range and M83 possesses different physical conditions and star formation environments, these differences likely reflect the combined influence of metallicity, ionization structure, nebular geometry, stellar population, and local ISM conditions on the ionic balance. Therefore, while the correlations identified here appear to be robust for the H\,\textsc{ii} regions in M33 and are broadly consistent with those in NGC~6822, the present comparison does not support their interpretation as universal relations applicable to all star-forming galaxies. A larger, homogeneous sample of spatially resolved JWST observations spanning a wider range of metallicities, radiation-field strengths, and nebular environments will be required to determine whether these ionization diagnostics represent universal characteristics of H\,\textsc{ii} regions or are instead dependent on the physical conditions of individual galaxies.

\section{Summary} 
\label{section: summary}

In this study, we present a spectroscopic analysis of five H\,\textsc{ii} regions in M33 using archival \textit{JWST}/MIRI observations. The high sensitivity and angular resolution of \textit{JWST} allow us to characterize a diverse set of mid-infrared tracers, including PAH bands, H$_2$ rotational transitions, hydrogen recombination lines, and ionic fine-structure lines. From these tracers, we derive key physical and ionization properties, including the H$_2$ rotational temperature, molecular hydrogen column density and mass, ionizing photon flux, ionization index, and ionization parameter.

We confirm that the $7.7/11.2$ PAH flux ratio decreases toward smaller physical scales, in agreement with the trend reported by \citet{maragkoudakis2018pahs}. The PAH $7.7/11.2$ flux ratio shows no clear correlation with radiation hardness, consistent with the results of \citet{maragkoudakis2018pahs}. Unlike their study, however, we do not find a positive correlation between the $7.7/11.2$ and $8.6/11.2$ ratios, while a moderate negative correlation is observed between $7.7/11.2$ and metallicity. These differences may be influenced by our small sample size ($N=5$). In contrast, the PAH-to-VSG ratio decreases with increasing radiation hardness, consistent with \citet{maragkoudakis2018pahs}.

We find that the PAH population is predominantly ionized, with an average cationic fraction of $\sim75\%$ for an absorption energy of 6.2 eV, despite the relatively moderate radiation fields indicated by the [Ne\,\textsc{iii}]/[Ne\,\textsc{ii}] ratios. The absence of a one-to-one relation between the ionization parameter and the $7.7/11.2~\mu$m ratio further indicates that PAH ionization depends on multiple factors, including PAH properties and local environmental conditions. The ionic abundances also show species-dependent positive and negative correlations with radiation hardness. Overall, while several trends are consistent with previous \textit{Spitzer}-based results, the limited sample size prevents statistically robust conclusions. A larger and homogeneous sample of M33 H\,\textsc{ii} regions observed with \textit{JWST}/MIRI will be essential for establishing the statistical significance of these trends and better constraining the interplay between radiation fields, PAHs, molecular gas, and ionized gas.

\begin{acknowledgments}

We thank the referee for the insightful and constructive comments, which have helped us improve the clarity and scientific presentation of the manuscript. We gratefully acknowledge the use of observations obtained with the JWST, retrieved from MAST. These data were acquired under Proposal ID~4297 (PI: Rogers, Noah Sidney James). JWST is operated by the National Aeronautics and Space Administration (NASA) in partnership with the European Space Agency (ESA) and the Canadian Space Agency (CSA). We also acknowledge the use of publicly available \textit{GALEX} near-ultraviolet (NUV) data, accessed through NASA’s SkyView Virtual Observatory, which contributed significantly to this study. This work was supported by the Indian Institute of Astrophysics (IIA) under the Department of Science and Technology (DST), Government of India.

\end{acknowledgments}





%
\facilities{JWST, GALEX}

\software{astropy \citep{2013A&A...558A..33A,2018AJ....156..123A,2022ApJ...935..167A}, JWST Data Analysis Tools (JDAT; SpecViz, CubeViz)}


\appendix

\renewcommand{\thefigure}{A\arabic{figure}}
\setcounter{figure}{0}

\section{PAH Spectra of the Remaining Four H II Regions}
\label{section: PAH Spectra of the Remaining Four H II Regions}

\begin{figure*}
\begin{center}
\resizebox{17.0cm}{6.0cm}{\includegraphics{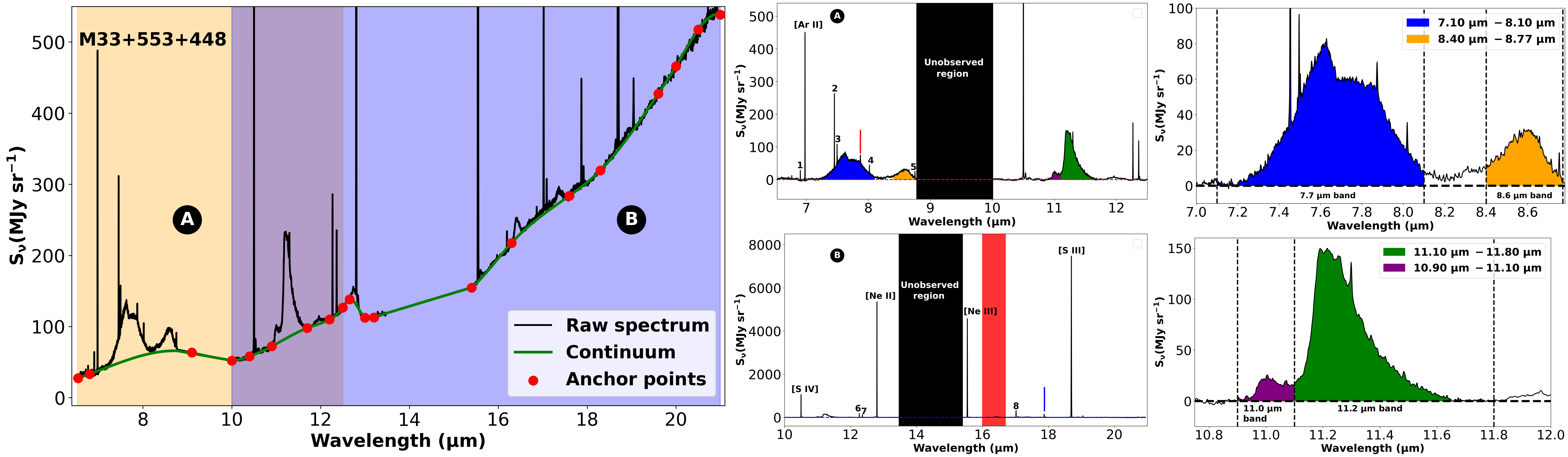}}
\caption{Same as Figure~\ref{Fig: M33-438+800 spectrum image}, but for the H\,\textsc{ii} region M33+553+448. The red vertical tick in the upper middle panel marks the expected position of a narrow emission feature that is not detected in the present analysis.}\label{Fig: M33+553+448 spectrum image}
\end{center}
\end{figure*}

\begin{figure*}
\begin{center}
\resizebox{17.0cm}{6.0cm}{\includegraphics{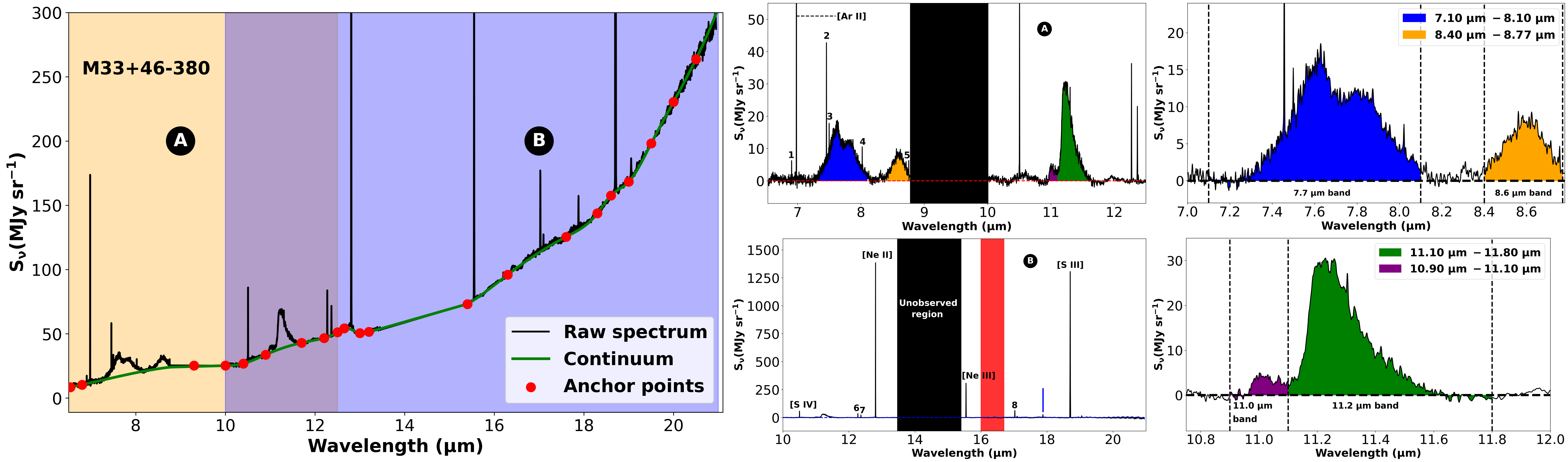}}
\caption{Same as Figure~\ref{Fig: M33-438+800 spectrum image}, but for the H\,\textsc{ii} region M33+46-380.}\label{Fig: M33+46-380 spectrum image}
\end{center}
\end{figure*}

\begin{figure*}
\begin{center}
\resizebox{17.0cm}{6.0cm}{\includegraphics{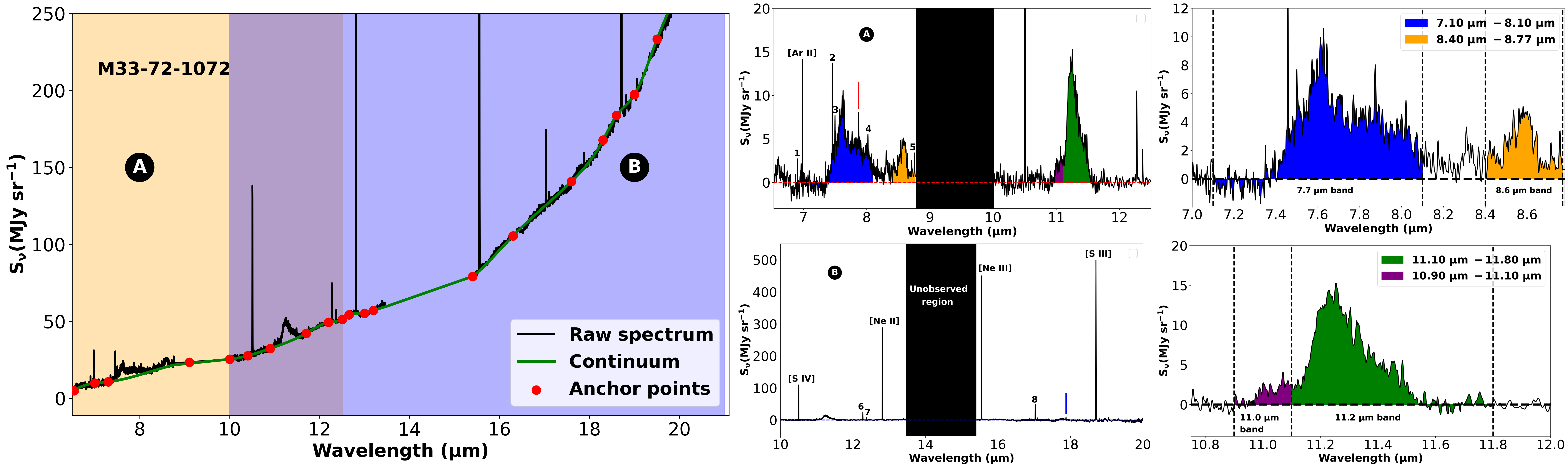}}
\caption{Same as Figure~\ref{Fig: M33-438+800 spectrum image}, but for the H\,\textsc{ii} region M33-72-1072 and the 16.4~$\mu$m PAH feature is not detected in this source. The red vertical tick in the upper middle panel indicates the expected position of a narrow emission feature that is not detected in the present analysis.}\label{Fig: M33-72-1072 spectrum image}
\end{center}
\end{figure*}

\begin{figure*}
\begin{center}
\resizebox{17.0cm}{6.0cm}{\includegraphics{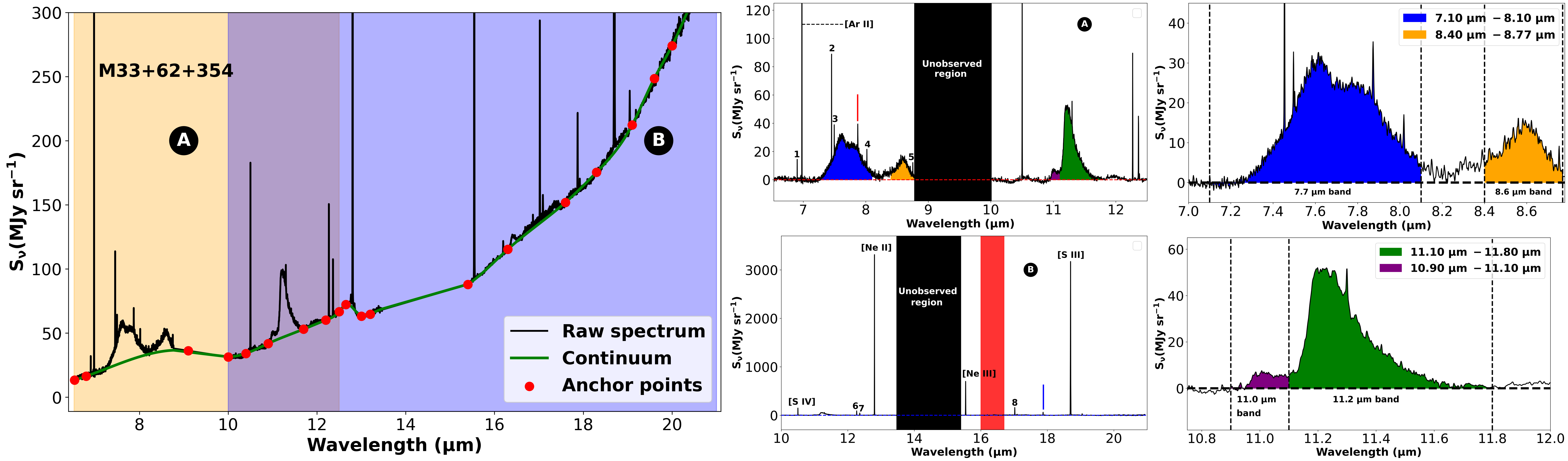}}
\caption{Same as Figure~\ref{Fig: M33-438+800 spectrum image}, but for the H\,\textsc{ii} region M33+62+354. The red vertical tick in the upper middle panel indicates the expected position of a narrow emission feature that is not detected in the present analysis.}\label{Fig: M33+62+354 spectrum image}
\end{center}
\end{figure*}

\renewcommand{\thefigure}{B\arabic{figure}}
\setcounter{figure}{0}

\section{Metallicity, H$_2$, and HI Spectra of the Remaining Four H II Regions}
\label{section: Metallicity, H2, and HI Spectra of the Remaining Four H II Regions}

\begin{figure*}
\begin{center}
\resizebox{17.0cm}{8.0cm}{\includegraphics{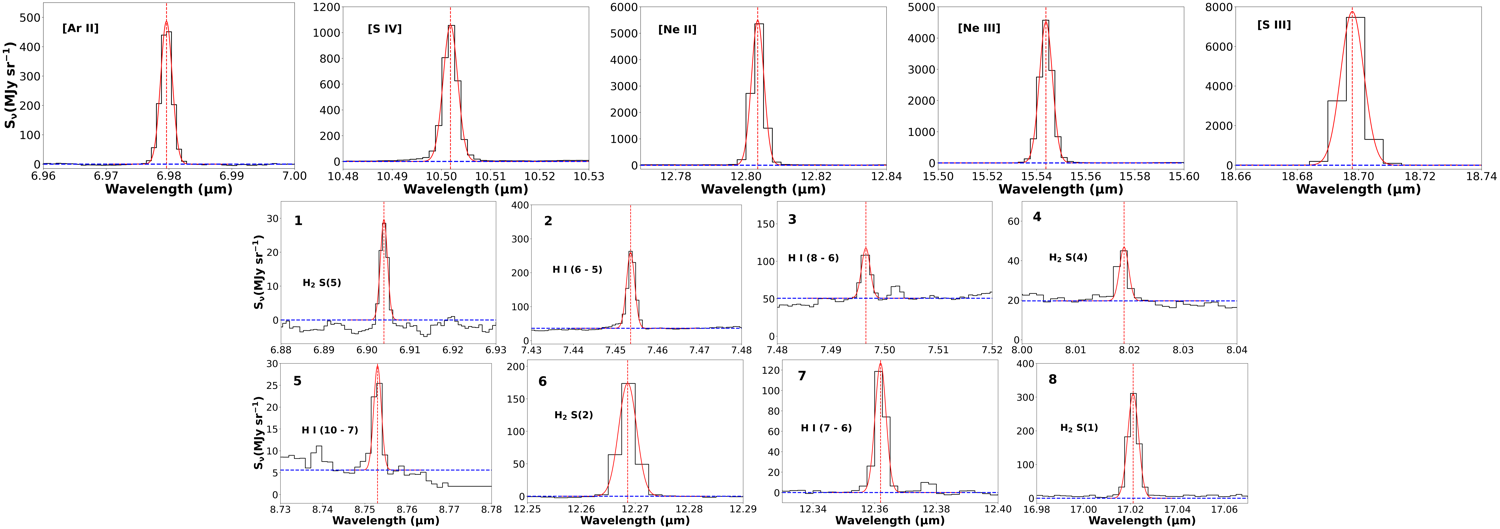}}
\caption{Same as Figure~\ref{Fig: M33-438+800 metallic spectrum}, but for the H\,\textsc{ii} region M33+553+448. The numerical labels corresponding to the hydrogen molecular rotational transitions and recombination lines follow the identification scheme defined in Figure~\ref{Fig: M33+553+448 spectrum image}.}\label{Fig: M33+553+448 metallic spectrum}
\end{center}
\end{figure*}

\begin{figure*}
\begin{center}
\resizebox{17.0cm}{8.0cm}{\includegraphics{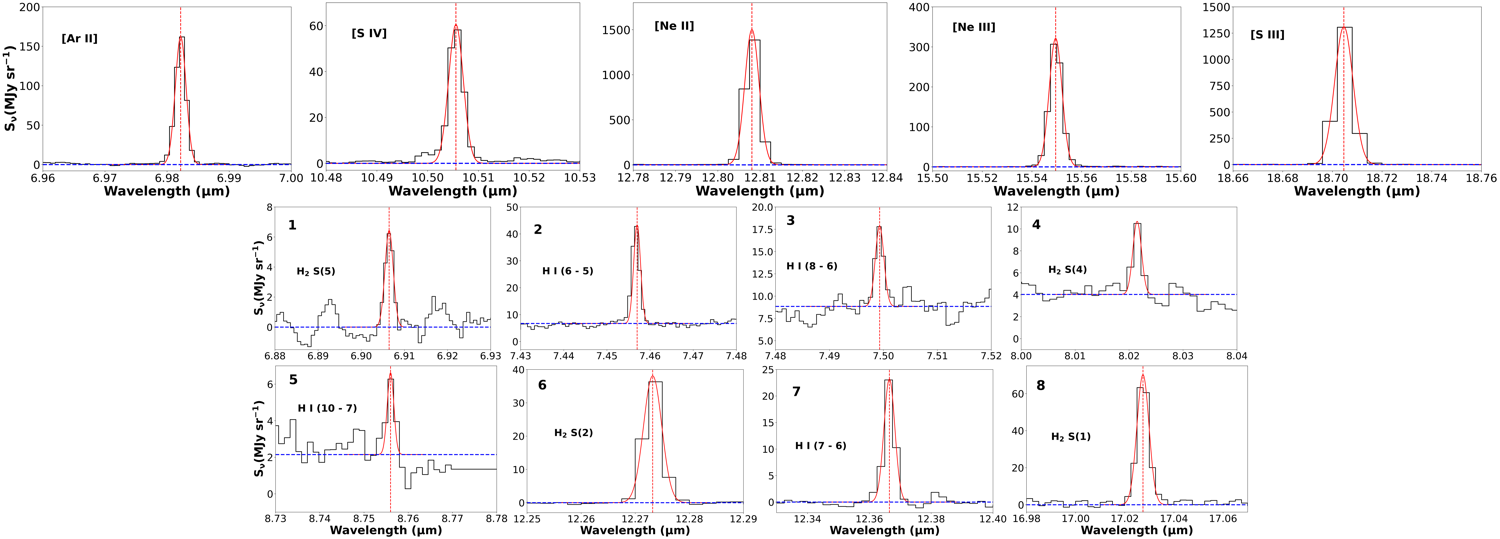}}
\caption{Same as Figure~\ref{Fig: M33-438+800 metallic spectrum}, but for the H\,\textsc{ii} region M33+46-380. The numerical labels corresponding to the hydrogen molecular rotational transitions and recombination lines follow the identification scheme defined in Figure~\ref{Fig: M33+46-380 spectrum image}.}\label{Fig: M33+46-380 metallic spectrum}
\end{center}
\end{figure*}

\begin{figure*}
\begin{center}
\resizebox{17.0cm}{8.0cm}{\includegraphics{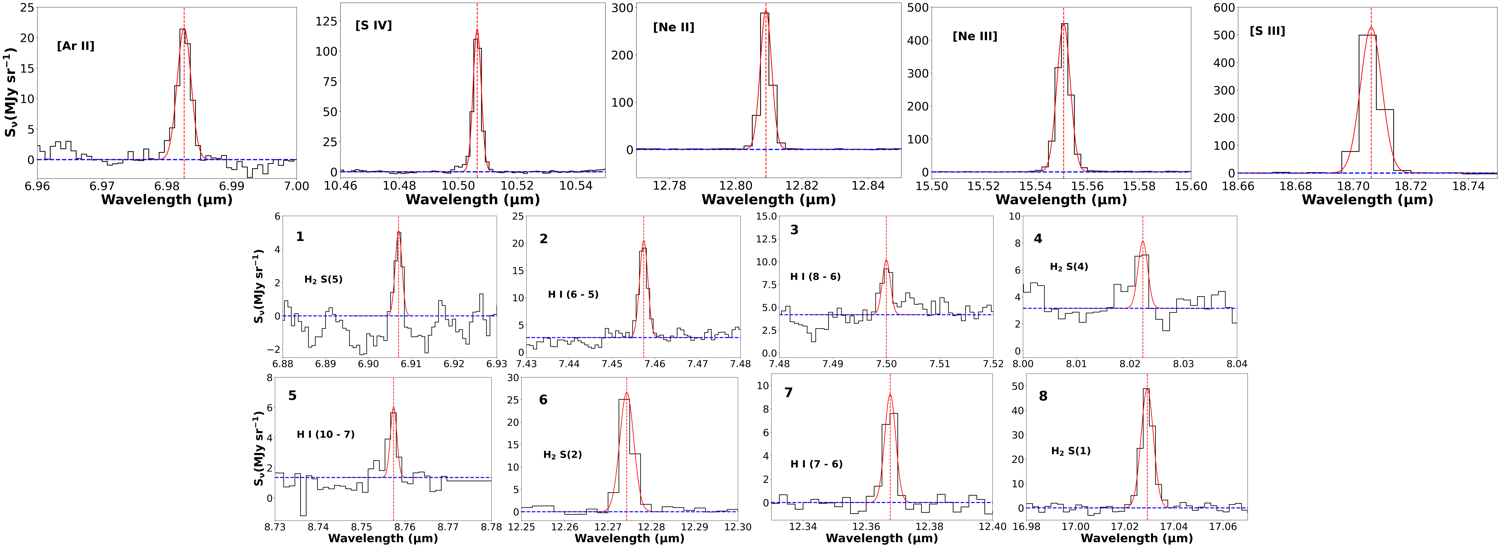}}
\caption{Same as Figure~\ref{Fig: M33-438+800 metallic spectrum}, but for the H\,\textsc{ii} region M33-72-1072. The numerical labels corresponding to the hydrogen molecular rotational transitions and recombination lines follow the identification scheme defined in Figure~\ref{Fig: M33-72-1072 spectrum image}.}\label{Fig: M33-72-1072 metallic spectrum}
\end{center}
\end{figure*}

\begin{figure*}
\begin{center}
\resizebox{17.0cm}{8.0cm}{\includegraphics{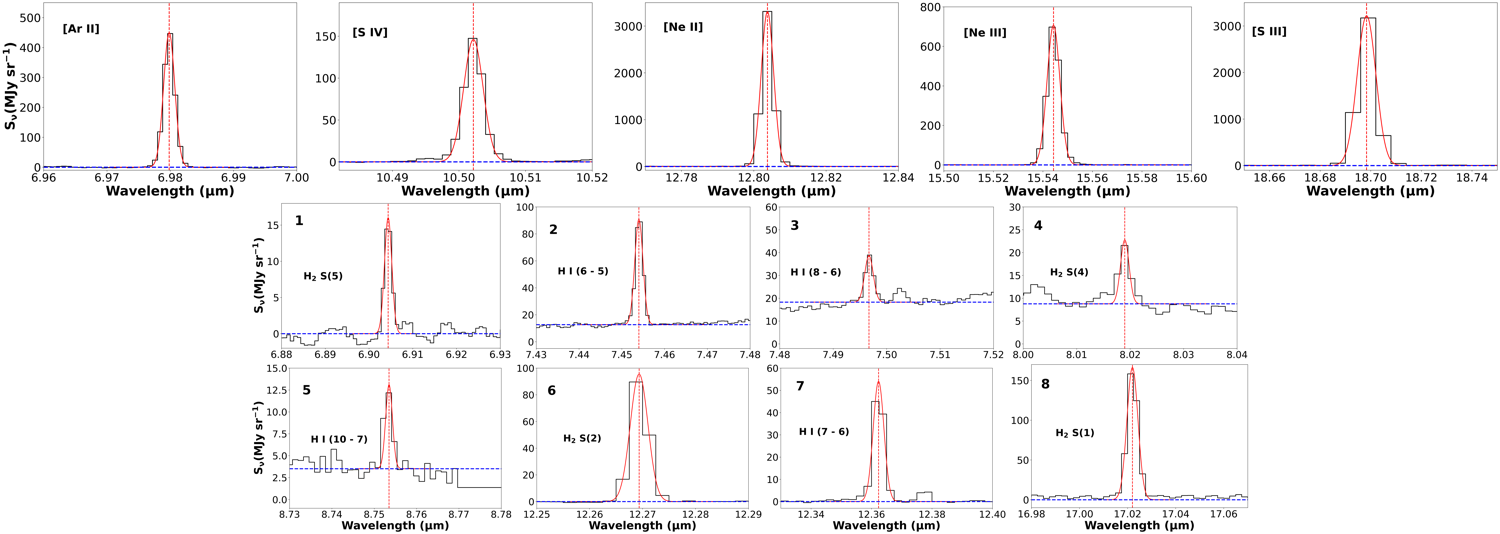}}
\caption{Same as Figure~\ref{Fig: M33-438+800 metallic spectrum}, but for the H\,\textsc{ii} region M33+62+354. The numerical labels corresponding to the hydrogen molecular rotational transitions and recombination lines follow the identification scheme defined in Figure~\ref{Fig: M33+62+354 spectrum image}.}\label{Fig: M33+62+354 metallic spectrum}
\end{center}
\end{figure*}



\bibliography{sample701}{}
\bibliographystyle{aasjournalv7}



\end{document}